\documentclass[reprint, amsmath, amssymb, floatfix, aps, prr,nofootinbib,longbibliography]{revtex4-2}

\usepackage{bm}
\usepackage{graphicx}
\usepackage{hyperref}
\usepackage{float}
\usepackage{xcolor}
\usepackage{fontawesome} 
\usepackage[ruled, vlined]{algorithm2e}
\usepackage{qcircuit}
\usepackage[caption=false]{subfig}
\usepackage{lineno}

\newcommand{\github}[1]{\href{#1}{\faGithubSquare}}
\newcommand{\esrgithub}{\github{https://github.com/MatthieuSarkis/Long-range-intermolecular-interactions-on-a-continuous-variable-quantum-computer}}

\graphicspath{{./}}

\begin{document}

\preprint{}

\title{Modeling Non-Covalent Interatomic Interactions on a Photonic Quantum Computer}

\author{Matthieu Sarkis}
\email{matthieu.sarkis@uni.lu}
\affiliation{
Department of Physics and Materials Science, University of Luxembourg, L-1511 Luxembourg City, Luxembourg
}

\author{Alessio Fallani}
\email{alessio.fallani@uni.lu}
\affiliation{
Department of Physics and Materials Science, University of Luxembourg, L-1511 Luxembourg City, Luxembourg
}

\author{Alexandre Tkatchenko}
\email{alexandre.tkatchenko@uni.lu}
\affiliation{
Department of Physics and Materials Science, University of Luxembourg, L-1511 Luxembourg City, Luxembourg
}

\date{\today}

\begin{abstract}
Non-covalent interactions are a key ingredient to determine the structure, stability, and dynamics of materials, molecules, and biological complexes. However, accurately capturing these interactions is a complex quantum many-body problem, with no efficient solution available on classical computers. A widely used model to accurately and efficiently model non-covalent interactions is the Coulomb-coupled quantum Drude oscillator (cQDO) many-body Hamiltonian, for which no exact solution is known. We show that the cQDO model lends itself naturally to simulation on a photonic quantum computer, and we calculate the binding energy curve of diatomic systems by leveraging Xanadu's \textsc{\textsc{strawberry fields}} photonics library. Our study substantially extends the applicability of quantum computing to atomistic modeling,
by showing a proof-of-concept application to non-covalent interactions, beyond the standard electronic-structure problem of small molecules. Remarkably, we find that two coupled bosonic QDOs exhibit a stable bond. In addition, our study suggests efficient functional forms for cQDO wavefunctions that can be optimized on classical computers, and capture the bonded-to-noncovalent transition for increasing interatomic distances.

\end{abstract}


\maketitle


\paragraph*{Introduction}

\begin{figure}
        \includegraphics[scale=0.095]{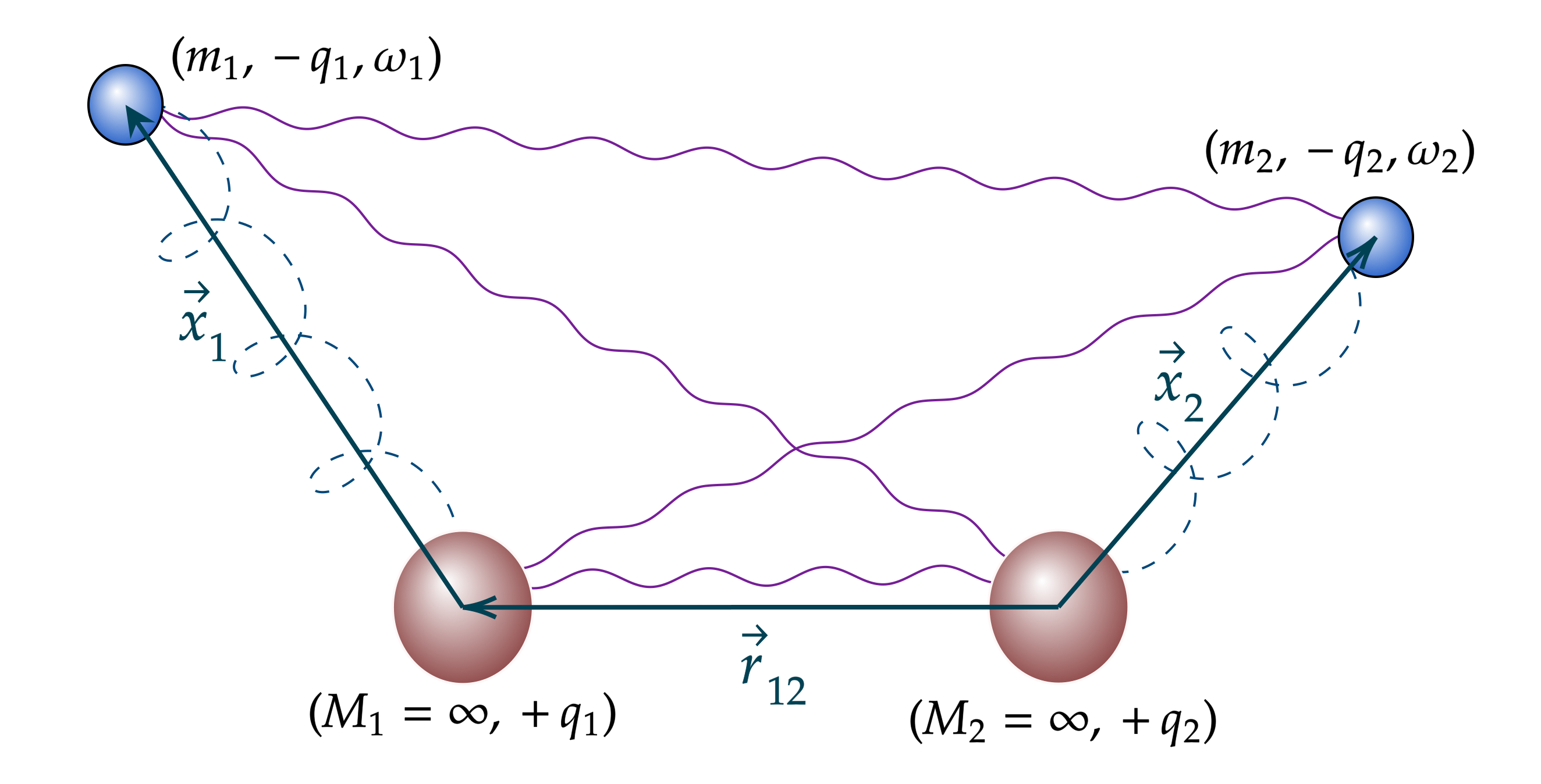}
        \caption{\label{fig:qdos}Illustration of a system composed of a pair of QDOs. The nuclei are considered to be infinitely massive. The drudons interact with their nucleus through a harmonic potential, but with the other QDO through Coulomb interaction. $\bm x_i$ denote the relative position of the drudon with respect to its nucleus in $\text{QDO}_i$. $\bm r_{12}$ denotes the position of $\text{QDO}_1$ with respect to $\text{QDO}_2$.}
    \end{figure}

    Materials and chemical modeling are considered to be among the most important applications of quantum computing. So far, most quantum computing algorithms have been applied to study short-range chemical bonds, primarily in small molecules, cf. \cite{cao2019quantum} for a review on the topic. However, long-range non-covalent interactions are key to understand many properties of molecules and materials \cite{hermann2017first, stohr2019theory}, motivating the developement of efficient and accurate models capturing these effects. Long-range forces originate from the electromagnetic interaction between electrically neutral atoms or molecules \cite{margenau2013theory,kaplan2006intermolecular,stone2013theory,hirschfelder2009intermolecular} and arise from the coupling of matter to the background quantum electrodynamic gauge field \cite{casimir1948influence,buhmann2013dispersion,buhmann2007dispersion,compagno1995atom,passante2018dispersion, cohen1997photons,cohen1998atom,bookpreparata,salam2009molecular,craig1998t}, and are at the core of the description of the properties of materials and macromolecules such as their structure \cite{hoja2019reliable}, stability \cite{hoja2018first,mortazavi2018structure}, dynamics \cite{stohr2019quantum,reilly2014role,PhysRevResearch.5.L012028} and coupling to a background gauge field \cite{kleshchonok2018tailoring, ambrosetti2022optical, Karimpour_2022}. The inclusion of dispersion interactions can be done in an economical way by means of many-body methods \cite{richardson1975dispersion,mahanty1973dispersion,RevModPhys.88.045003,tkatchenko2015current,ren2012random,harl2009accurate,dobson2012calculation,parsegian2005van,becke2006simple,becke2006exchange,grimme2010consistent,grimme2006semiempirical,tkatchenko2012accurate,massa2021many}, in particular through the so-called many-body dispersion (MBD) framework, whose accuracy was proven in the literature \cite{tkatchenko2012accurate, ambrosetti2014long}. In the MBD framework, drawing inspiration from the Drude atomic model, the response of valence electrons in atoms is assumed to be linear and this can be implemented through the introduction of quantum Drude oscillators (QDO), for which a harmonic potential is assumed between an effective electron (called \textit{drudon} in this context) and an oppositely-charged nucleus, as illustrated in Fig. \ref{fig:qdos}. QDOs define a compact coarse-grained quantum-mechanical model for dispersion forces, in which molecules are then defined as a collection of QDOs interacting through \textit{dipole-dipole interaction} \cite{doi:10.1063/1.1743992}. Though simple, this system was shown to capture long-range phenomena even in large biomolecular systems \cite{https://doi.org/10.48550/arxiv.2205.11549}. By construction, the MBD framework relies on the dipole-dipole approximation to the Coulomb interaction, and therefore comes with the usual limitations of multipolar-type expansions, leaving aside any contribution coming from higher-order couplings. These limitations can be addressed via multipolar generalizations of the pairwise second-order perturbative approaches \cite{massa2021beyond,massa2021many,becke2006simple,becke2006exchange}.
    We propose here, in the spirit of the Full Configuration Interaction (FCI) approach of \cite{sadhukhan2016quantum}, to study the MBD model with full Coulomb interaction between its constituents. Going beyong the quadratic dipolar QDO Hamiltonian of course comes at the cost of loosing integrability of the model, and this is where numerical methods, and in particular quantum computing approches become relevant.

    Here, we show that quantum computing can be applied to efficiently and accurately model non-covalent interactions. Our study substantially extends the applications of quantum computing to atomistic modeling, by probing the use of Noisy Intermediate Scale Quantum (NISQ) algorithms to the study of concrete quantum chemistry models, beyond the standard electronic-structure problem of small molecules \cite{peruzzo2014variational, kandala2017hardware, nam2020ground}. 
    
    We remark that the authors of Ref.~\cite{anderson2022coarse} implemented a Variational Quantum Eigensolver for the simulation of an effective one-dimensional QDO model. There, the quantum harmonic oscillator Hilbert space, built as a Fock space, was truncated at some fixed level $\Lambda$, and mapped to the Hilbert space of a system of $\lceil\log_2\Lambda\rceil$ qubits. The Hamiltonian was then mapped to a linear combination of Pauli strings. They studied there in great detail the case of a linear coupling between the QDOs, and explored the case of a quartic potential; However it is known on the one hand that the quadratic model has a known analytical solution, and we argued on the other hand that it is beneficial to explore QDO models beyond the multipolar expansion. Moreover the drudons were allowed to move either \textit{longitudinally} or \textit{transversally} with respect to the axis connecting the QDOs. However, as we will show, both of these models/configurations do not capture the most critical properties of the full-fledged three-dimensional cQDO model, namely existence of a bound state, and smoothness of the binding curve \cite{sadhukhan2016quantum, ditte2023molecules}. The present work adresses both limitations. On the other hand, our work also differs from a choice of hardware point of view. Indeed, in our case we choose the photonic-based continuous-variable quantum computing paradigm \cite{lloyd1999quantum, pati2000quantum, pati2003deutsch, braunstein2005quantum, andersen2010continuous, lomonaco2002continuous, zhong2020quantum, tillmann2013experimental}, guided by the intrinsic bosonic nature of QDOs, therefore effectively mapping a bosonic problem onto a bosonic hardware. The Fock space of a single harmonic oscillator is directly identified with the Fock space of one mode (one photon channel, in quantum optics language) of the quantum electrodynamic gauge field. In particular, the position and momentum of the drudon particle are directly identified with the position and momentum quadratures of the electromagnetic field, and a collection of $N$ QDOs is then represented by a collection of $3N$ photon channels in an optical circuit. Similar ideas were successfully used to simulate bosonic Euclidean quantum fields \cite{marshall2015quantum, yeter2022quantum} on a lattice. Concerning the quantum simulation of interacting harmonic oscillators in a first quantized framework, let us mention \cite{Culver:2021rxo}, where the authors encode simple supersymmetric quantum mechanical systems into a system of qubits, using again a hard cutoff for the bosonic degrees of freedom. In our case, and following  \cite{schuld2021machine, killoran2019continuous, arrazola2019machine, enomoto2022continuous, peruzzo2014variational}, an optical circuit composed of parameterized linear gates like beam-splitters, phase shifters, but also Gaussian gates implementing displacement and squeezing, and quartic gates (Kerr gates) is optimized through a Variational Quantum Eigensolver (VQE) type algorithm in order to accurately prepare the ground state wavefunction of the cQDO model. For the implementation, we leverage on Xanadu's \textsc{strawberry fields} quantum machine learning framework for continuous-variable hybrid quantum-classical optimization \cite{schuld2021machine, killoran2019continuous, arrazola2019machine, bromley2020applications}. Following \cite{arrazola2019machine} for state preparation, we probe our approach using \textsc{strawberry fields} simulator API. We successfully derive the binding energy curve for a pair of QDOs, and study various properties of the corresponding ground state wavefunction. We note in particular the existence of a bound state, as already showed in \cite{sadhukhan2016quantum, ditte2023molecules}, and we observe an interesting behavior of the quantum mutual information as a function of the interatomic distance, in correlation with the formation of the bound state as the two QDOs approach each other.
    
\paragraph*{Definition of the model}

    The Hamiltonian describing a system of $N$ Quantum Drude Oscillators in three dimensions is given by:
    \begin{linenomath}\begin{equation}
        H=\sum_{i=1}^N\left[\frac{\bm{p} _i^2}{2m_i} + \frac{1}{2}m_i\omega_i^2\bm{x} _i^2\right] +\sum_{i<j}V_\text{Coul}\left(\bm{x} _i, \bm{x} _j\right)\,,
    \end{equation}\end{linenomath}
    with the Coulomb interaction containing contributions from interacting drudon-drudon and drudon-nucleus pairs:
    \begin{linenomath}\begin{equation}
    \begin{split}
        \frac{V_\text{Coul}\left(\bm{x} _i, \bm{x} _j\right)}{q_iq_j}=&\ \frac{1}{r_{ij}} - \frac{1}{|\bm{r}_{ij}
        + \bm{x} _i|} - \frac{1}{|\bm{r}_{ij}  - \bm{x} _j|} \\
        & + \frac{1}{|\bm{r}_{ij} - \bm{x} _j + \bm{x} _i|}\,,
    \end{split}
    \end{equation}\end{linenomath}
    and where $\bm x_i$ denotes the relative position of drudon $i$ with respect to its nucleus, and $\bm r_{ij} = \bm r_i - \bm r_j$ denotes the position of $\text{nucleus}_i$ with respect to $\text{nucleus}_j$. The usual approach consists in solving the theory in the multipolar expansion framework, in which the potential can be expressed as a power series in the inverse distance separating the two nuclei:
    \begin{linenomath}\begin{equation}
        V_\text{Coul}\left(\bm{x} _i, \bm{x} _j\right)= \sum_{n\geq 0} V_n\left(\bm{x} _i, \bm{x} _j\right)\,,
    \end{equation}\end{linenomath}
    with the following scaling behavior $V_n\left(\bm{x} _i, \bm{x} _j\right)\propto r_{ij}^{-n-3}$.
    The potential $V_0$ corresponds then to the dipole-dipole interaction, and is at the core of the Many Body Dispersion (MBD) model. $V_1$ corresponds to the dipole-quadrupole interaction, and $V_2$ to the quadrupole-quadrupole and dipole-octupole interaction.

    One obvious limitation of the multipolar expansion, which assumes a large distance between the nuclei with respect to the typical separation between the drudons and their nucleus, is the lower bound it imposes on the interatomic distance. One can easily see that within the MBD (dipole-dipole) model by direct diagonalization of the quadratic Hamiltonian in terms of normal modes. In that case, one of the normal modes (the center of mass mode) develops a purely imaginary frequency at short range. Higher order physical effects are also neglected in the MBD model, motivating the study of the QDO model with full Coulomb interaction potential between its constituents.

    Let us define the following dimensionless position and momentum operators associated to QDO $i$:
    \begin{linenomath}\begin{equation}
        \bm{X}_i := \sqrt{\frac{m_i\omega_i}{\hbar}}\,\bm{x}_i\,,\ \ \ \ \ \bm{P}_i := \frac{\bm{p}_i}{\sqrt{\hbar m_i\omega_i}}\,,
    \end{equation}\end{linenomath}
    in terms of which the Hamiltonian reads
    \begin{linenomath}\begin{equation}
    \begin{split}
        H =&\ \sum_{i=1}^N\frac{\hbar\omega_i}{2}\left(\bm X_{i}^2 + \bm P_{i}^2\right) \\
        & + \sum_{i<j}V_\text{Coul}\left(\sqrt{\frac{\hbar}{m_i\omega_i}}\bm{X} _i, \sqrt{\frac{\hbar}{m_j\omega_j}}\bm{X} _j\right)\,.
    \end{split}
    \end{equation}\end{linenomath}
    One can define the $3N$ creation and annihilation operators
    \begin{linenomath}\begin{equation}
        \bm a_{i} = \frac{\bm X_{i} + i\bm P_{i}}{\sqrt 2}\,,\ \ \ \ \ \bm a^\dagger_{i} = \frac{\bm X_{i} - i\bm P_{i}}{\sqrt 2}\,,
    \end{equation}\end{linenomath}
    in terms of which the Hamiltonian reads
    \begin{linenomath}\begin{equation}
    \begin{split}
        &H = \sum_{i=1}^N\hbar\omega_i\left(\bm a_{i}^\dagger\cdot\bm a_{i} +\frac{3}{2}\right) \\
        & + \sum_{i<j}V_\text{Coul}\left(\sqrt{\frac{\hbar}{m_i\omega_i}}\frac{\bm a_i + \bm a_i^\dagger}{\sqrt 2}, \sqrt{\frac{\hbar}{m_j\omega_j}}\frac{\bm a_j + \bm a_j^\dagger}{\sqrt 2}\right)\,.
    \end{split}
    \end{equation}\end{linenomath}
    Let us from now on restrict the problem to a pair of QDOs ($N=2$) for concreteness, though the following developements carry to systems of $N$ coupled QDOs. We denote by $d$ the interatomic distance.

    In order to reduce the complexity of the problem, let us define one-dimensional instances of the QDO model as follows: we restrict the movement of the two drudons to be along a common axis (directed by a unit vector $\hat{\bm e}_\theta$) which form an angle $\theta\in[0,\pi/2]$ with respect to the vector $\bm r_{12}$ connecting the two nuclei. We therefore have a family of one-dimensional models which can be obtained from the full-fledged 3d model simply by setting to zero the contribution from the oscillator modes belonging to the plane perpendicular to $\hat{\bm e}_\theta$. Let us denote by $(X_i, P_i)_{i\in\{1,2\}}$ the remaining position and momentum degree of freedoms. As limiting cases, we obtain models in which the drudons are constrained to move either in the direction parallel to the axis separating the two nuclei ($\theta=0$), or perpendicular to the latter ($\theta=\pi/2$). Those two models were studied in \cite{anderson2022coarse} in the dipole-dipole approximation. As mentioned in the introduction, in that paper the authors encode the states in the truncated Fock space of the system into the state of a set of qubits, and run VQE-type algorithms on IBQ quantum processors. However as we will see, the angle $\theta$ captures the competition between \textit{existence of binding} (for small $\theta$) and \textit{smoothness} (for large $\theta$), and interesting one-dimensional models actually sit at values of $\theta$ in the open segment $(0,\pi/2)$.

    In the case of a generic angle $\theta$ and interatomic distance $d$, the one-dimensional Coulomb potential reads:
    \begin{linenomath}\begin{equation}
    \begin{split}
        &\frac{V_\text{Coul}^{\theta, d}(x_1, x_2)}{q_1q_2} = \frac{1}{d} - \frac{1}{\sqrt{d^2+2d(\cos\theta) x_1+x_1^2}} \\
        &- \frac{1}{\sqrt{d^2-2d(\cos\theta) x_2+x_2^2}} \\
        &+\frac{1}{\sqrt{d^2 - 2d(\cos\theta) (x_2-x_1) + (x_2-x_1)^2}}\,.
    \end{split}
    \end{equation}\end{linenomath}
    We therefore have a family of Hamiltonians parameterized by $(\theta, d)\in[0, \pi/2]\times\mathbb R_{>0}$:
    \begin{equation}
    \label{eq:finalHamiltonian}
    \begin{split}
        H_{\theta, d}=&\ \frac{\hbar\omega_1}{2}\left(P_1^2 + X_1^2\right) + \frac{\hbar\omega_2}{2}\left(P_2^2 + X_2^2\right)\\
        &+V_\text{Coul}^{\theta, d}\left(\sqrt{\frac{\hbar}{m_1\omega_1}}X_1, \sqrt{\frac{\hbar}{m_2\omega_2}}X_2\right)\,.
    \end{split}
    \end{equation}
    Alternatively, in terms of creation and annihilation operators, one has
    \begin{equation}
    \label{eq:Hamiltonian_heterodyne}
    \begin{split}
        H_{\theta, d} &= \hbar\omega_1\left( a_{1}^\dagger a_{1} +\frac{1}{2}\right) + \hbar\omega_2\left(a_{2}^\dagger a_{2} +\frac{1}{2}\right) \\
        & + V_\text{Coul}^{\theta, d}\left(\sqrt{\frac{\hbar}{m_1\omega_1}}\frac{a_1 + a_1^\dagger}{\sqrt 2}, \sqrt{\frac{\hbar}{m_2\omega_2}}\frac{a_2 + a_2^\dagger}{\sqrt 2}\right)\,.
    \end{split}
    \end{equation}
    For the numerical simulations, we set $\hbar=4\pi\epsilon_0=1$ as well as $m_i=q_i=\omega_i=1$ for both QDOs.\newline

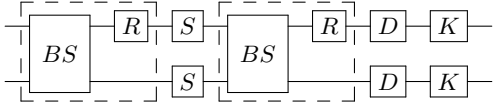
\begin{figure}
        \begin{linenomath}\begin{equation*}
            \Qcircuit  @C=1em @R=1em {
            & \multigate{1}{BS}  & \gate{R}  & \gate{S}  & \multigate{1}{BS}  & \gate{R}  & \gate{D}  & \gate{K}  & \qw \gategroup{1}{5}{2}{6}{.7em}{--} \\
            & \ghost{BS}  & \qw  & \gate{S}  & \ghost{BS}  & \qw  & \gate{D}  & \gate{K} &
         \qw \gategroup{1}{2}{2}{3}{.7em}{--}
         \\
        }
    \end{equation*}\end{linenomath}
    \caption{\label{fig:quantum_circuit}One layer in the optical quantum circuit. The various gates, namely the beamsplitters, rotation, squeezing, displacement and Kerr gates are collectively parameterized by $\omega$.}
    \end{figure}

\paragraph*{Photonic circuit and Variational Algorithm}
    From a molecular physics and long-range intermolecular perspective, we are mainly interested in knowing the ground state of the system (\ref{eq:finalHamiltonian}). Among the various approaches, the hybrid quantum-classical variational algorithms were shown to be particularly efficient at capturing the properties of potentially complicated quantum states. Following \cite{killoran2019continuous, arrazola2019machine}, we apply a continuous-variable version of the VQE algorithm to extract the ground state of the QDO system. The ground state is obtained by optimizing the parameters of a parameterized optical circuit composed of linear multi-modes gates (two-mode beamsplitters and rotation gates), Gaussian gates (single-mode squeezing and displacement gates), as well as non-Gaussian gates (Kerr gates) implementing non-linearity. 
    A generic multimode Gaussian state can be decomposed into a sequence of two-mode beam-splitters and single mode rotations, squeezing and displacement. Inserting a non-Gaussian operation ('non-linearities' in the classical machine learning language) in between each Gaussian transformation, allows to evade the strictly Gaussian realm, and provide enough flexibility to capture arbitrarily complicated wavefunctions by statcking multiple layers, increasing the expressivity of the ansatz space. As discussed later, one can justify a posteriori this specific choice of non-linearity by observing that our ground state is well-captured by cat-like states, whose preparation is known to require the use of Kerr interactions. Within a layer, the number of parameters in the model scales quadratically in the number of photon channels (due to the beam-splitters). We refer the reader to fig. \ref{fig:quantum_circuit} for an illustration of a single layer composing the architecture of the quantum neural network.
    In the procedure, the relative coordinate of the drudons with respect to their nucleus is directly identified with the position quadrature of the quantized electromagnetic field along the two channels of the optical quantum circuit.

    Overall, denoting by $\omega$ the set of all parameters, the circuit implements a unitary transformation $U(\omega)$ acting on an input reference state that we simply take to be the Fock vacuum state $|0\rangle\otimes|0\rangle$. The state prepared by the circuit is therefore given by $|\psi(\omega)\rangle = U(\omega)|0\rangle\otimes|0\rangle\,.$
    Once the ansatz state $|\psi(\omega)\rangle$ has been produced, one extracts the value of the energy in that state, given by $\langle\psi(\omega)|H|\psi(\omega)\rangle\,.$
    Let us denote expectations in the state $|\psi(\omega)\rangle$ by angular brackets $\langle\cdot\rangle$.
    To be specific, let us take our model (\ref{eq:finalHamiltonian}), and let us denote by angular brackets the expectation in state $|\psi(\omega)\rangle$. One has
    \begin{equation}
    \label{eq:loss}
    \begin{split}
        \langle H\rangle =&\ \hbar\omega_1\left(\langle a_{1}^\dagger a_{1}\rangle+\frac{1}{2}\right)+\hbar\omega_2\left(\langle a_{2}^\dagger a_{2}\rangle+\frac{1}{2}\right)\\
        &+ \left\langle V_\text{Coul}^{\theta, d}\left(\sqrt{\frac{\hbar}{m_1\omega_1}} X_1, \sqrt{\frac{\hbar}{m_2\omega_2}} X_2\right)\right\rangle
    \end{split}
    \end{equation}
    by linearity of the expectation. On the second line one has to compute an expression of the form $\langle f(X_1, X_2)\rangle$. One therefore needs to extract the joint statistics of the position quadratures by preparation and measurements of the state $|\psi(\omega)\rangle$ in the quadrature basis, as summarized in alg. (\ref{alg:statistics_computation}). Once the joint probability density $\rho$ of $(X_1, X_2)$ in the state $|\psi(\omega)\rangle$ is known, one can compute
    \begin{linenomath}\begin{equation}
        \langle f(X_1, X_2)\rangle = \int_{\mathbb R^{6}}f(x_1, x_2)\rho(x_1, x_2)\,\text{d}x_1\text{d}x_2\,,
    \end{equation}\end{linenomath}
    where the integral above should be understood as a finite sum over a sufficiently refined grid $G_X\times G_X$ in the position quadratures plane. For the numerical simulation we take a linear grid $G_X$ composed of 500 points in the interval $[-6,6]$.

    Let us note that alternatively, and following \cite{arrazola2019machine}, the simulator of \textsc{strawberry fields} actually provides direct access to the output statevector expressed in the Fock basis:
    \begin{linenomath}\begin{equation}
        |\psi(\omega)\rangle = \sum_{n_1,n_2=0}^\infty \alpha_{n_1n_2}(\omega)|n_1\rangle\otimes|n_2\rangle\,.
    \end{equation}\end{linenomath}
    The Fock space is actually truncated at a some fixed energy level, we choose that cutoff level to be 5. This choice of cutoff is not a real limitation and a larger value could be chosen instead, but 5 is sufficient, as was observed in particular using a FCI approach \cite{sadhukhan2016quantum}.
    The amplitude of a specific pair of the quadratures $(X_1, X_2)$ is then given by:
    \begin{linenomath}\begin{equation}
        \langle X_1,X_2|\psi(\omega)\rangle = \sum_{n_1,n_2=0}^\infty \alpha_{n_1n_2}(\omega)\prod_{i=1}^{2}\frac{e^{-\frac{X_i^2}{2}}H_{n_i}(X_i)}{\sqrt{\pi^{1/2}2^{n_i}n_i!}}\,,
    \end{equation}\end{linenomath}
    in terms of the Hermite polynomials. The joint law of the quadratures in the state $|\psi(\omega)\rangle$ is then given by
    \begin{linenomath}\begin{equation}
        \rho(X_1,X_{2}) = \left|\langle X_1,X_2|\psi(\omega)\rangle\right|^2\,.
    \end{equation}\end{linenomath}
    After extracting as well the mean photon numbers $\langle a_1^\dagger a_1\rangle$ and $\langle a_2^\dagger a_2\rangle$, one obtains $\langle H\rangle$.

    We then define the cost function
    \begin{equation}
    \label{eq:definition_cost}
        \mathcal C(\omega) := \langle\psi(\omega)|H|\psi(\omega)\rangle\,,
    \end{equation}
    and update the parameters $\omega$ of the optical circuit in order to minimize that cost. This is summarized in alg. (\ref{alg:training}).
    \begin{algorithm}
        \caption{Extract distribution of position quadratures}\label{alg:statistics_computation}
            \textbf{Parameters:} statevector $|\psi\rangle$, finite subset $\mathcal {G}\in\mathbb R$, shots $M\in\mathbb N$

            \KwResult{Probability distribution of $(X_1, X_2)$ in state $|\psi\rangle$ discretized over the grid $G_X\times G_X$}\

            \For{$m=1$ to $|\mathcal G|$}{
                Initialize $N_m \gets 0$;
            }
            \For{$j=1$ to $M$}{
            Measure the state $|\psi\rangle$ in the discretized position quadratures basis, obtain the eigenvalue $X_m$, set $N_m\gets N_m+1$\;
            }
            \For{$m=1$ to $|\mathcal G|$}{
                Normalize $N_m \gets N_m / M$;
            }
            \textbf{return} $\{N_m\}_{m=1}^M$.
    \end{algorithm}

    \begin{algorithm}
        \caption{Training of the parameterized photonic circuit}\label{alg:training}
        \textbf{Parameters:} Model $(\theta, d)$, $N_\text{steps}\in\mathbb N$, initial circuit parameters $\omega_0\in\mathbb R^K$, learning rate $\eta\in\mathbb R_+$

        \KwResult{Optimized hyperparameters $\omega\in\mathbb R^K$}\

        Initialize hyperparameters $\omega \gets \omega_0$\;
        \For{$i=1$ to $N_\text{steps}$}{
        Compute the loss $\mathcal C$ according to eq. (\ref{eq:definition_cost})\;
        Compute the gradient $\nabla_\omega\mathcal C$ with the shift rule\;
            Update the parameters $\omega \gets \omega - \eta\nabla_\omega\mathcal C$\;
        \textbf{end for}
        }
        \textbf{return} $\omega$.
    \end{algorithm}
    As a side remark, let us note that given the form of the Hamiltonian in terms of creation and annihilation operators (\ref{eq:Hamiltonian_heterodyne}), an alternative to the measurement of the position quadratures and photon number operators could be to perform a measurement in the coherent basis through heterodyne measurements.\newline

\paragraph*{Binding energy curve}

    We gather here the results of the simulations. We focus on the case of 2 QDOs. In particular we study the profile of the binding energy as a function of the distance between the two nuclei, and make a few observation about the behavior of the entanglement entropy of the system.

\begin{figure}[!ht]
        \centering
        \subfloat{
        \hspace*{-.45cm}
          \includegraphics[scale = 0.9]{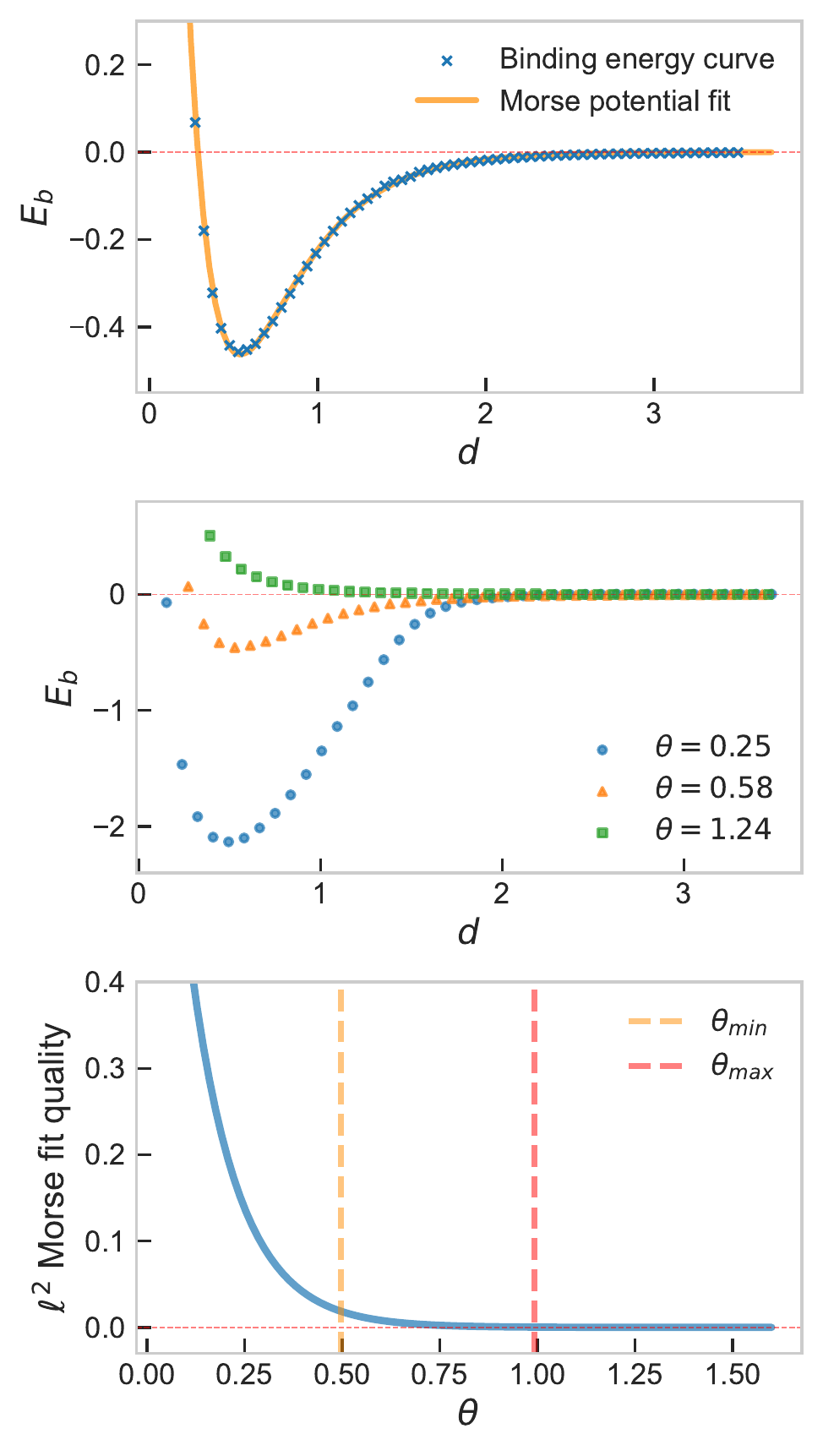}
        }
        \caption{\label{fig:binding}\textbf{On the top.} Binding energy curve for the model at angle $\theta=0.58$, together with its Morse fit. \textbf{On the middle.} Binding energy curve for three different values of the angle $\theta$ illustrating the tension between models close to $\theta=0$ and models close to $\theta=\pi/2$. For small angle (blue curve),  the strong curvature prevents a good Morse fit, while for large angle (green curve), the transverse configuration of the drudons prevents formation of a bound state. The orange curve corresponds an intermediate angle exhibiting both binding and an excellent Morse fit. \textbf{On the bottom.} \textbf{Blue curve:} Quality of the Morse fit as a function of the angle $\theta$, defined as the $\ell^2$-norm of the difference between the simulation and the Morse fit. \textbf{Yellow line:} Minimal angle above which the models are considered smooth. \textbf{Red line:} Maximal angle above which the models do not exhibit binding anymore.}
    \end{figure}

    We fix a grid $G_\theta\times G_d\subset [0, \pi/2]\times (0, 3.5]$, with $\text{card}(G_\theta)=20$ and $\text{card}(G_d)=200$. For each pair $(\theta, d)$ in the grid we perform the continuous-variables VQE algorithm to extract properties of the ground state of the Hamiltonian (\ref{eq:finalHamiltonian}). Let us denote by $|\psi_{\theta, d}\rangle$ the corresponding converged ground state.
    The binding energy is defined as the difference between the ground state energy of the system of interacting QDOs and the ground state energy of a system of uninteracting QDOs $H_0$, i.e. with electric charge turned off:
    \begin{equation}
        E^\theta_b(d) = \langle\psi_{\theta, d}|H_{\theta, d}|\psi_{\theta, d}\rangle - \langle\psi_0|H_0|\psi_0\rangle
    \end{equation}
    In fig. \ref{fig:binding} we report the binding energy curve, namely the value of the binding energy as a function of the interatomic distance $d$, for a fixed value of the angle $\theta$. We choose $\theta=0.58$ to illustrate most of the results.

    On fig. \ref{fig:binding} (top) we observe a perfect agreement with a fit of Morse type \cite{herzberg1945molecular, apanavicius2021morse, le2006accurate, roy2007new, chiumorse2010}:
    \begin{equation}
    \label{eq:morse_fit}
        f(d) = E_b\left(e^{-2\frac{d-d_b}{s}} - 2e^{-\frac{d-d_b}{s}}\right)\,.
    \end{equation}
    The location of the bound state is given by $d_b\simeq 0.54$, with energy $-E_b\simeq 0.46$. The length scale $s$ is given by $s\simeq 2.75$. The quality of the Morse fit actully increases monotonically with the angle, cf. fig. \ref{fig:binding} (bottom), the quality being defined by the $\ell^2$-distance between the numerical result and the fit.

    By varying the value of the angle $\theta$, we observe two different regimes. For small values of the angle, the curve deviates more and more from a Morse curve, and in the extreme longitudinal case $\theta=0$, becomes highly non-smooth at short interatomic distances. This small angle regime is also characterized by the existence of a negative global minimum of the binding energy, hence by the existence of a bound state. On the other hand, as the angle increases, the $\ell^2$-quality of the fit becomes better. However, passed some value of the angle, the global minimum disappears together with the corresponding bound state. This competition between smoothness and existence of a bound state is illustrated in fig. \ref{fig:binding} (middle), and of course signals the presence of a phase transition in the model where $\theta$ plays the role of control parameter. Below a critical value $\theta_\star$ of the order parameter, the system is characterized by the existence of a bound state, which disappear above the critical value of $\theta_\star$.

    Both regimes can be understood physically as follows: for very small angle $\theta\ll \theta_\star$, and in particular for the longitudinal model $\theta=0$, as the two QDOs are getting closer and closer to each other, unstable configurations in which the two drudons are getting arbitrarily close to each other start appearing. Indeed, space being 1-dimensional and the two drudons moving along a common axis, as the two drudons get close to one another, the associated Coulomb repulsion component of the ground state energy diverges. On the other hand, for large angle $\theta\gg \theta_\star$, and \textit{a fortiori} for the transverse model $\theta=\pi/2$, two main configurations of the drudons may occur (thinking classically) depending on the relative position of the drudons with respect to the axis connecting the two nuclei. In the first configuration, the drudons are sitting on opposite sides. In that case, the dominant contribution to the energy between the two QDOs is the Coulomb repulsion between the nuclei. In the second configuration, the drudons are on the same side. In that case, in addition to the repulsive force between the two nuclei, one can also add up the repulsion force between the two drudons, leading to an even more repulsive scenario. Summing up, no binding can occur at $\theta=\pi/2$, and by smoothness of the binding energy as a function of $\theta$, this should also be the case in an open neighborhood of $\theta=\pi/2$.

    The above observation suggests the following recipe. The longitudinal model predicts the existence of negative minima of the binding energy curve. It is however unstable due to the configurations of superposed drudons, as explained above. One can then \textit{regularize} this 1d model by allowing for a non-zero angle $\theta$, and slightly increase it until reaching a certain level of smoothness, that we have chosen here to be quantified by the quality of a Morse fit. The angle should however not be too large, smaller than the transition point beyond which the bound state disappears. This procedure defines a small range of models characterized by an angle $\theta\in[\theta_\text{min}, \theta_\text{max}]$, as illustrated on fig. \ref{fig:binding} (bottom). Finally, the QDO parameters usually represent the properties of real atoms, such as their polarizability and dispersion coefficients. Following for instance the paper of Jones et al. \cite{Jones2010QuantumDO}, the parameters could be fixed by fitting the dispersion coefficients predicted by the cQDO model to the reference (computed \textit{ab initio} or measured experimentally) dispersion coefficients of the underlying atomic species. In addition, recently an optimized QDO parameterization has been proposed using only well-known dipolar properties of gas-phase atoms~\cite{doi:10.1021/acs.jpclett.3c01221}.

\paragraph*{Phase space representation}

\begin{figure*}
        \includegraphics[scale=0.75]{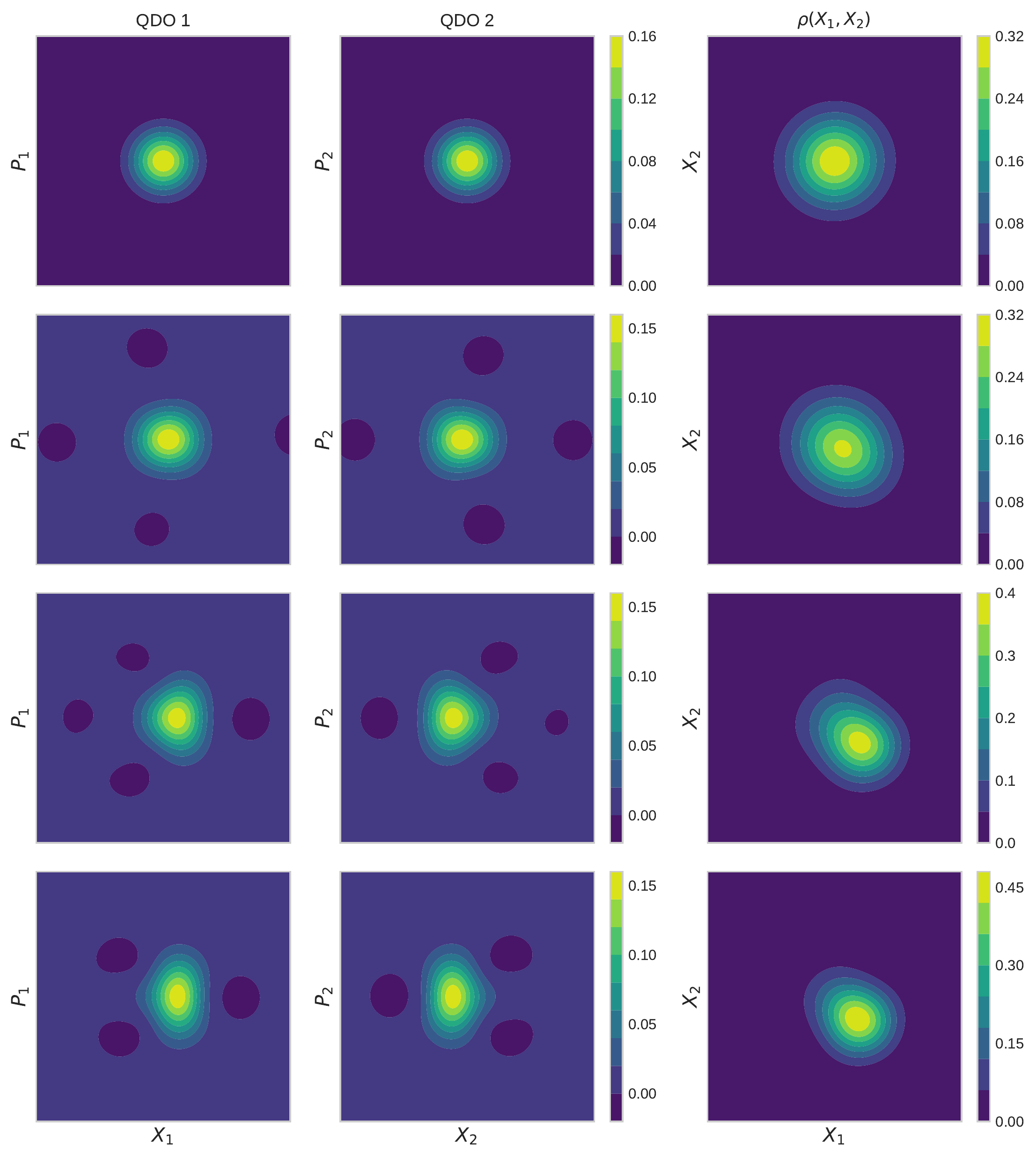}
        \caption{\label{fig:wigners_joint}\textbf{From top to bottom:} $d=3.16$: QDOs are far apart, $d=1.36$: QDOs start feeling each other, $d=0.82$: entanglement entropy is maximal, $d=0.54$: deep in the bound state. \textbf{From left to right:} Wigner distribution of the left QDO, Wigner distribution of the right QDO, joint position quadrature distribution of the two drudons.}
    \end{figure*}

\begin{figure*}
        \includegraphics[scale=0.85]{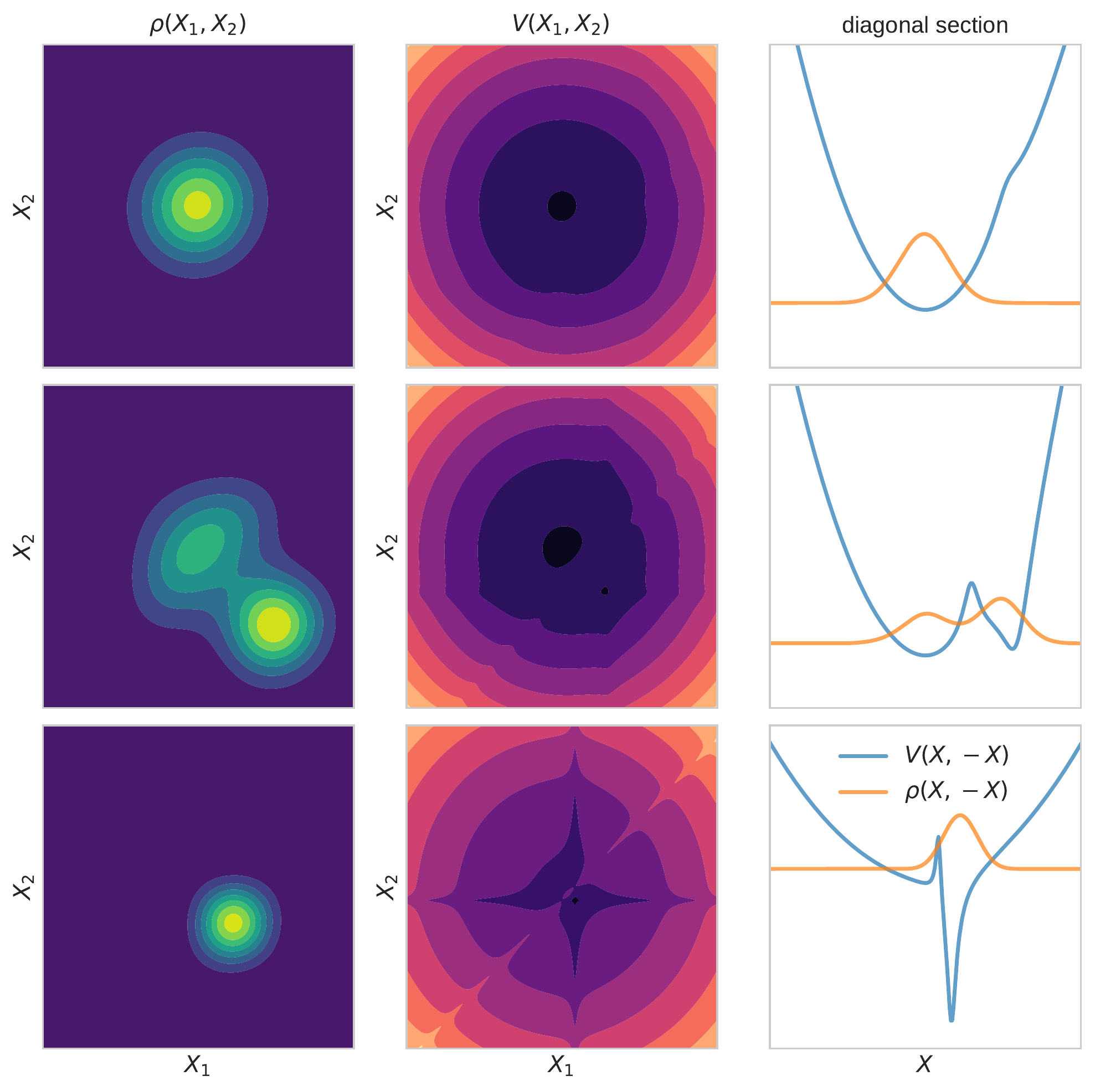}
        \caption{\label{fig:classical_potential_small_angle}\textbf{From top to bottom:} $d=3.16$: QDOs are far apart, $d=1.75$: Maximal interaction between the QDOs, $d=0.51$: deep in the bound state. \textbf{From left to right:} joint position quadrature distribution of the two drudons, classical potential energy $V(X_1,X_2) = V_\text{Coul}(X_1,X_2) + (X_1^2 + X_2^2) /2$, diagonal section of the joint distribution and of the potential. Tunneling between the quadratic and Coulomb wells is maximal at the intermediate interatomic distance as can be seen from the bimodal joint distribution, and the neat appearance of two local vacua in the classical potential. This figure corresponds to the small angle $\theta=0.17$.}
    \end{figure*}

    In order to have a better intuition about the nature of the ground state of the system, let us turn to its representation in the quadratures phase space. In fig. \ref{fig:wigners_joint}, we provide two representations of the system at different values of the interatomic distance (for the model at angle $\theta=0.58$).
    On the left side, from top to bottom, we represent the marginal Wigner function of each of the two QDOs for closer and closer distances. At large distance $d=3.16$, the two QDOs are far apart and do not feel each other. This can be seen by the fact that the marginal Wigner functions are characteristic of purely Gaussian states. At distance $d=1.36$, the two QDOs start feeling each other's presence. This is illustrated by the two marginal Wigner functions starting to be elongated, mostly along the position quadrature axis. Distance $d=0.82$ corresponds to the distance of maximal entanglement entropy, as discussed a bit later in the paper. We observe in that case that the marginal Wigner functions are characterized by a significant spread in both position and momentum quadrature directions. Finally, at the bound state distance $d=0.54$, the marginal Wigner functions appear to be mostly elongated along the momentum quadrature axis. This is clear since even though the configuration of the system has reached stability, characterized by the fact that the Wigner functions suggest a fairly neat localization of the drudons in space, the drudons being close to each other lead to an increase in the spread of the momentum density. Observe also the appearance of regions of negativity of the partial Wigner functions, signature of the non-classicality of the state of the system \cite{Chabaud:2021pnh}, and witness of the entanglement in bipartite states \cite{arkhipov2018negativity}. The right column of fig. \ref{fig:wigners_joint} instead illustrates the joint probability distribution of the position quadratures. We observe again a Gaussian behavior at large distances, as it should for two independent harmonic oscillators, a maximal elongation at the maximal entanglement entropy point, and a more localized appearance at the bound state distance. Let us mention that a refined analysis around the distance of maximal position quadrature density spread shows that the unimodal joint position quadrature probability distribution develops a bimodal profile, with maximum of the original mode located around the origin (both drudons being in expectation centered at the locus of their respective nucleus), and maximum of the newly appeared mode shifted away from the origin. This bimodal feature of the joint position quadrature density, signature of tunneling, is very easily observable for a smaller angle theta, as can be seen on fig. \ref{fig:classical_potential_small_angle}, which is consistent with the large curvature of the binding curve for such small angles, at the corresponding range of interatomic distances. After tunneling, the mode initially centered at the origin disappears, and remains only the shifted mode describing a stable bound state configuration in which each of the QDOs has developed a non-zero expected dipole moment, oriented in opposite directions.\newline
    
\paragraph*{Ground state wavefunction and entanglement entropy}

\begin{figure}[h]
        \centering
        \hspace*{-.45cm}
        \includegraphics[scale = 0.9]{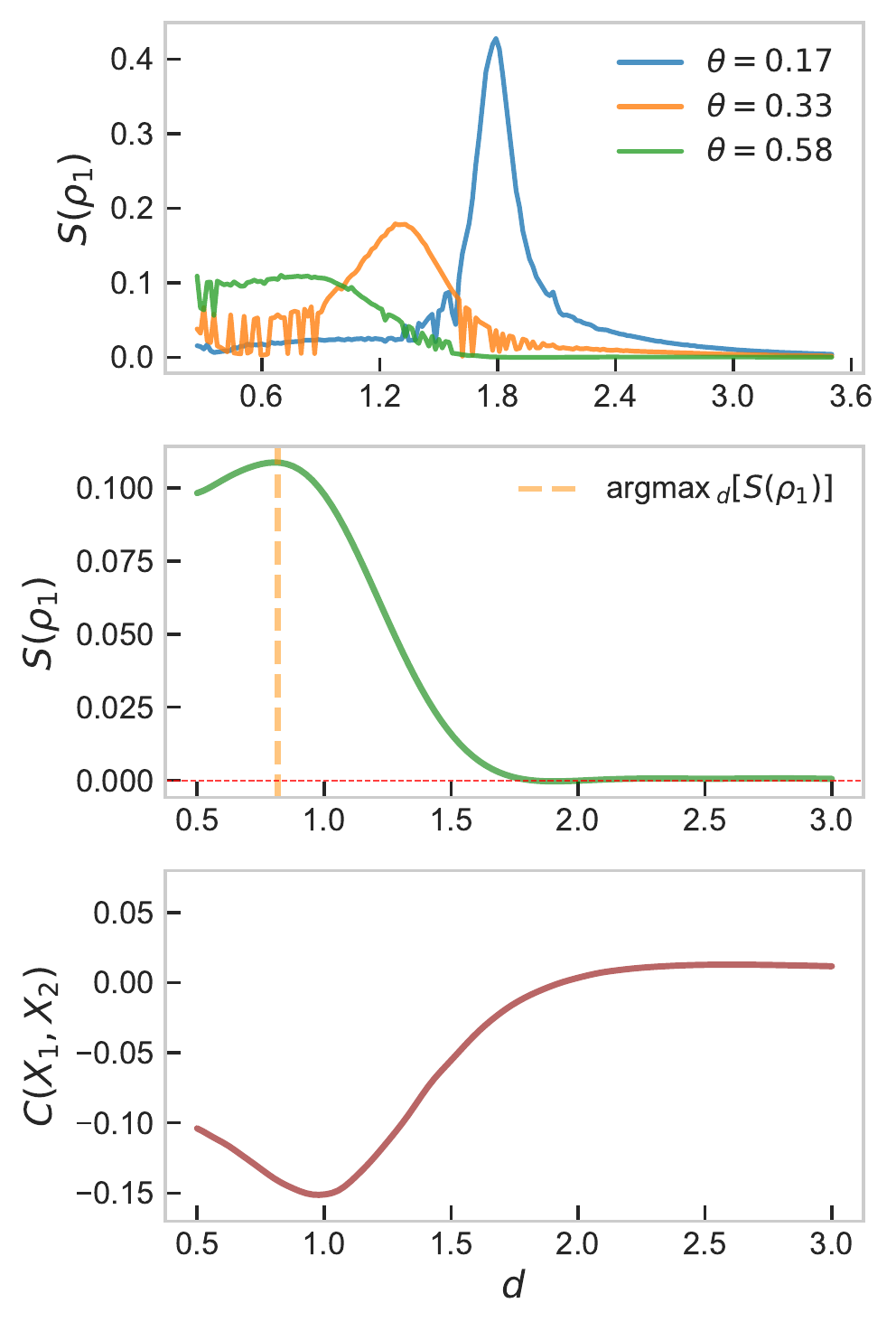}%
    \caption{\label{fig:entropy_correlation}\textbf{On the top.} Entanglement entropy as a function of the distance $d$ between QDOs for multiple values of $\theta$. \textbf{On the middle.} Smoothing of the entanglement entropy vs. interatomic distance at angle $\theta=0.58$. \textbf{On the bottom.} Position quadratures correlation coefficient vs. interatomic distance at angle $\theta=0.58$.}
    \end{figure}

    In order to gain more intuition about the information possessed by one part of our bipartite system concerning the other part, let us introduce the partial density matrix associated to $\text{QDO}_1$. The total state of the system, we recall, is expressed in the Fock basis as the pure state:
    \begin{linenomath}\begin{equation}
        |\psi\rangle = \sum_{n_1,n_{2}=0}^\infty \alpha_{n_1n_2}|n_1\rangle\otimes|n_2\rangle\,.
    \end{equation}\end{linenomath}
    This state is directly accessible in \textsc{strawberry fields} when using the simulator, and could be obtained on a genuine hardware by state tomography techniques \cite{Lvovsky:2009zz}. Note that the full knowledge of the state is not necessary for the optimization of the VQE parameters. We use this knowledge here solely to analyze the physics of the obtained ground state. Given the pure state describing the full system, the density matrix of the system is simply given by:
    \begin{linenomath}\begin{equation}
        \rho = \sum_{\substack{n_1,n_2 \\ m_1,m_2}} \alpha^*_{m_1m_2}\alpha_{n_1n_2}|n_1\rangle\langle m_1|\otimes|n_2\rangle\langle m_2|\,.
    \end{equation}\end{linenomath}
    The partial trace associated to $\text{QDO}_1$ is therefore given by:
    \begin{linenomath}\begin{equation}
        \rho_1 = \sum_{n, m, l} \alpha^*_{ml}\alpha_{nl}\,|n\rangle\langle m|\,.
    \end{equation}\end{linenomath}
    Since the state of the total system is pure, $\text{QDO}_2$ can be interpreted as purifying the system composed solely of $\text{QDO}_1$. The two QDOs therefore have identical von Neumann entropy $S(\rho_1)$, the entanglement entropy. The quantum mutual information of the system is therefore given by
    \begin{linenomath}\begin{equation}
        I(1:2) = S(\rho_1) + S(\rho_2) - S(\rho) = 2S(\rho_1) \,,
    \end{equation}\end{linenomath}
    with the von Neumann entropy being defined as
    \begin{linenomath}\begin{equation}
        S(\rho) = -\text{Tr}\left[\rho\log\rho\right]\,.
    \end{equation}\end{linenomath}
    The profile of the entanglement entropy for different values of the angle $\theta$ as a function of the interatomic distance is provided in fig. \ref{fig:entropy_correlation} (top).

    In this figure we can see how for small values of $\theta$ the entanglement entropy presents a sharp peak which gradually turns into a plateau-like behaviour as $\theta$ increases ($\theta = 0.58$ is the highest angle for which a peak is found). 
    
    As a comment, we observe fluctuations in the entanglement entropy, mostly for $\theta = 0.33$. One can check that the distance between the density matrices in the sense of the Frobenius norm between subsequent values of the interatomic distance shows fluctuations in the same windows as those observed in the entanglement entropy. The fluctuations are still relatively small, but suggest that there exists a small neighbourhood of acceptable ground states in the ansatz space. This in turn suggests that these fluctuations could be mitigated by imposing a stronger convergence criterium in the VQE training loop.
    
    A kernel fit of the entanglement entropy for angle $\theta = 0.58$ is provided in fig. \ref{fig:entropy_correlation} (middle).

    This finding, together with the behaviour of the joint probability reported in fig. \ref{fig:wigners_joint}, gives us an intuition of what happens to the ground state of the system along the binding process, allowing to \textit{open the VQE blackbox}. For $d\rightarrow \infty$ the two QDOs will not interact and hence the natural state for them will be the vacuum state $|0\rangle|0\rangle$ (ground state of independent harmonic oscillators). On the other hand when $d \sim d_{\mathrm{bonding}}$ we see that the system is shifted towards an antisymmetric configuration where $\langle X_1\rangle =-\langle X_2\rangle$ and $\langle P_1\rangle = \langle P_2\rangle = 0$, which can be approximated by the bipartite coherent state $|\alpha\rangle|-\alpha\rangle$ with $\alpha = \langle X_1\rangle$. In the transition region instead the system will pass through a strongly entangled state (hence the peak in von Neumann entropy), which is reflected in the position joint probability by the transition from a single mode to a bimodal distribution, as we discussed previously. This is naturally represented as a superposition of the form
    \begin{linenomath}\begin{equation}
        \frac{1}{\mathcal{N}}\left(|0\rangle\otimes|0\rangle + |\alpha\rangle\otimes|-\alpha\rangle\right),
    \end{equation}\end{linenomath}
    where $\mathcal{N}$ is a normalization factor, which can be viewed as a displaced Schr\"odinger cat state. For each value of the interatomic distance, one obtains the closest cat state to the ground state by optimizing the displaced cat state parameters through maximization of the state fidelity (overlap) between the two states.
    \begin{figure}[h]
        \centering
        \subfloat{%
          \includegraphics[clip,width=1.\columnwidth]{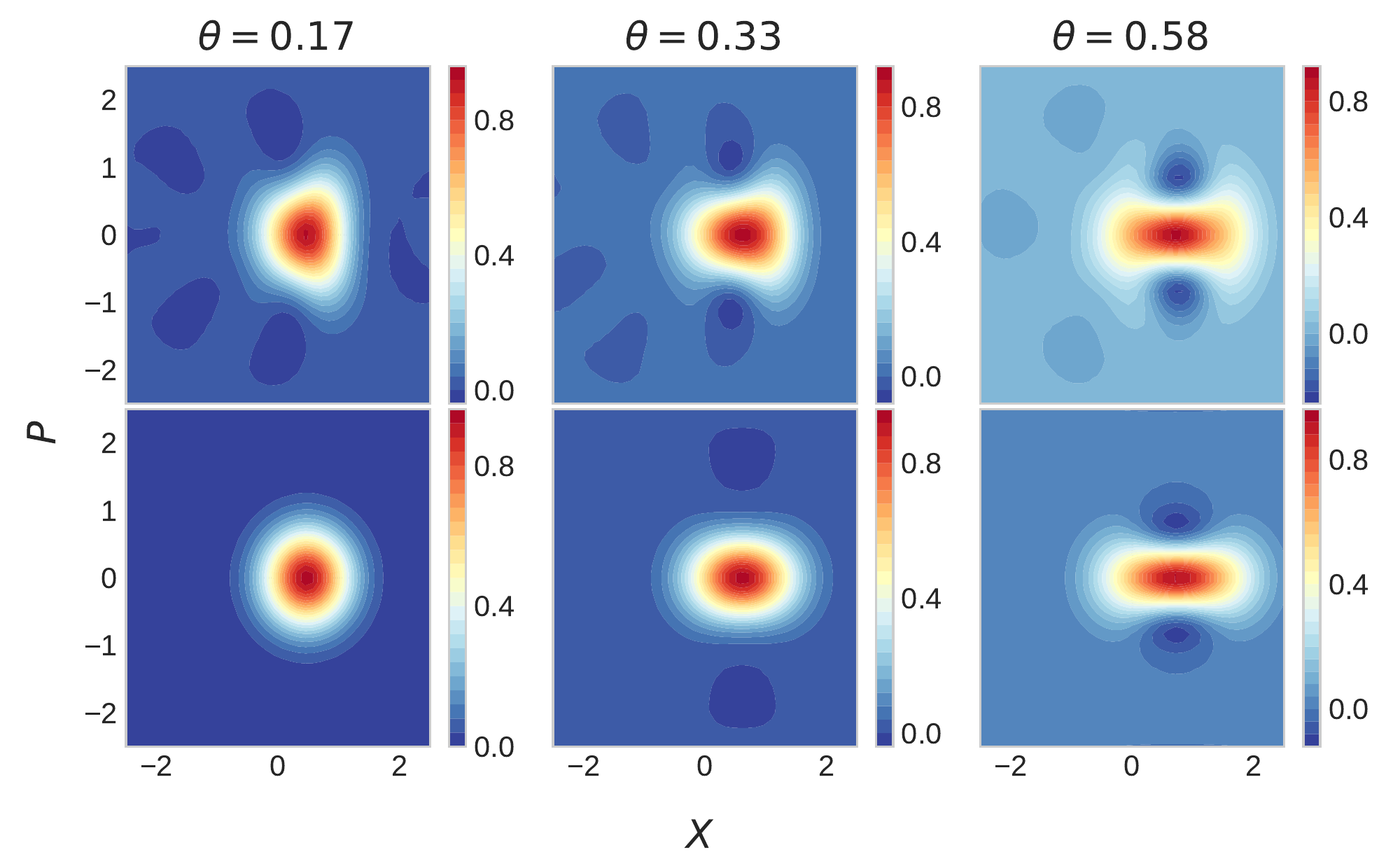}%
        }
        \caption{\label{fig:fidecats}Comparison between the states of the system on the entropy peaks (top) and the fitted ansatz in the same point (bottom). We plot the Wigner function sliced along the plane $(X = X_1 = -X_2, P = P_1 = -P_2$).}
    \end{figure}
    Although there is a slight deterioration in the fit for decreasing values of theta, this minimal ansatz can indeed approximate the ground state of the system up to a good degree for all the considered thetas: it manages to capture the bonding mechanism (states for which there is an entropy peak) with $\mathcal{F}\sim 0.96$. A comparison at the level of Wigner functions between the original and fitted states for those bonding points is reported in fig. \ref{fig:fidecats}. Let us mention that the production of cat states typically involves the use of Kerr and squeezing gates \cite{puri2017engineering, grimm2019kerr}, both of which being present in our quantum neural network architecture, Kerr gates being the source of non-Gaussianity and entanglement of the ansatz. A more detailed study concerning cat states in the context of long-range interactions will appear in future work.

    Given the joint position quadrature density, another interesting quantity to consider is the quantum correlation between the position quadrature of the two QDOs, which appears to be highly correlated to the behavior of the entanglement entropy, cf. fig. \ref{fig:entropy_correlation} (middle and bottom). Denoting again by angular brackets the expectation of an observable in the ground state, we define the correlation coefficient $C(X_1, X_2)$ by:
    \begin{linenomath}\begin{equation}
        C(X_1, X_2) = \frac{\langle X_1X_2\rangle - \langle X_1\rangle\langle X_2\rangle}{\sqrt{\langle X_1^2\rangle - \langle X_1\rangle^2}\sqrt{\langle X_2^2\rangle - \langle X_2\rangle^2}}\,.
    \end{equation}\end{linenomath}

    Even though we do not have a neat theoretical understanding of it, we note that the maximum of the entanglement entropy, as well as the minimum of the correlation coefficient, converges towards the inflexion point of the binding energy curve as the value of $\theta$ increasing. A zeroth order line of argument could be the following: the binding energy is composed of a Coulomb repulsion and an attractive correlation energy contribution:
    \begin{linenomath}\begin{equation}
        E(d)=E_\text{Rep}(d)+E_\text{Corr}(d)
    \end{equation}\end{linenomath}
    The inflexion point $d_\star$, which in terms of the Morse fit (\ref{eq:morse_fit}) is located at $d_\star = d_b + \log(2)s$,
    can be interpreted as the transition between a large distance regime in which it is beneficial for the system to increase correlation in order to lower its ground state energy, and a short range regime in which the Coulomb repulsion becomes dominant. Further understanding of this phenomenon in the language of quantum phase transitions would be very interesting, but is left for a more theoretical investigation. Let us mention that the use of cat states as discussed above, for which the expression of the quantum mutual information turns out to be particularly simple, could also provide an approximate understanding for the presence of the peak. This will be discussed further in future work.\newline

\paragraph*{Conclusion}

    In this work, we showed that continuous variables photonics quantum computing is particularly adapted to the study of the full-Coulomb cQDO model, the fundamental degrees of freedom of the latter being bosonic in nature. Beyond standing as a proof-of-concept that NISQ algorithms can be successfully applied to quantum chemistry problems beyond the usual approach using the second quantized formulation of the electronic structure problem for small molecules, we observed that the cQDO model for Many Body Dispersion interaction can be further simplified by reducing it to a single effective spatial dimension, while still capturing the existence of binding. The obtained groundstate wavefunction suggests that Schr\"odinger cat states can provide an economical ansatz space for systems exhibiting bonding-like behavior. This observation happens to also hold for the fully-fledged 3D cQDO model. Finally, note that Morse type binding curves are usually characteristic of covalent bonding in molecules and materials, suggesting that the cQDO model can potentially be used (up to a reparameterization of its defining parameters) to capture effects beyong those motivating its introduction, namely dispersion interactions. In particular the exponential repulsion behavior is reminiscent of exchange effects, which are a priori not built in the initial definition of the model. The extension of our study to many-body cQDO Hamiltonians are also relatively straightforward.

    The quantum neural network architecture used in our study is quite general, and one could wonder whether a more specific architecture tailored to the cQDO model could be engineered. However, the many-body nature of the model makes it difficult to guess any a priori correlations between the various photon channels. A possible approach to get a priori intuition about a possible entanglement graph among the various modes could be, drawing inspiration from the MBD model~\cite{tkatchenko2012accurate}, to first work in the dipole-dipole approximation. In this case the Hamiltonian is quadratic, hence quasi-free, and we can diagonalize it first to get the MBD normal modes. These normal modes do not constitute of course the eigenmodes of the full Coulomb model, but could provide a first coarse intuition about a possible circuit architecture tailored to the model. It would constitute an interesting direction, and we postpone that to possible later exploration.

    Finally, concerning the possibility of scaling up to very large systems, one is obviously confronted to the fact that the bigger the system, the higher dimensional the position quadrature grid. Drawing inspiration from the classical reinforcement learning literature, in which a similar Monte Carlo average of a random variable has to be obtained during the training loop, it could be interesting to see whether a very coarse estimate made out of very few  samples could be enough for the VQE training loop to converge to a sensible result.
    \newline

\paragraph*{Acknowledgments}

    We would like to thank Dahvyd Wing and Kyunghoon Han for helpful discussions. We thank Dmitry Fedorov and Matteo Barborini for carefully reading and commenting the draft of this article. A.F. aknowledges financial support from the European Union's Horizon 2020 research and innovation program under Marie Sklodowska-Curie grant agreement No 956832, "Advanced Machine learning for Innovative Drug Discovery" (AIDD). A.T. acknowledges funding via the FNR-CORE Grant ``BroadApp'' (FNR-CORE C20/MS/14769845) and ERC-AdG Grant ``FITMOL''.
    \newline

%

    
\paragraph*{Code Availability}

    The reader will find an open source python code accompanying this paper at \esrgithub.
    \newline

%

\appendix

%


\begin{thebibliography}{78}%
\makeatletter
\providecommand \@ifxundefined [1]{%
 \@ifx{#1\undefined}
}%
\providecommand \@ifnum [1]{%
 \ifnum #1\expandafter \@firstoftwo
 \else \expandafter \@secondoftwo
 \fi
}%
\providecommand \@ifx [1]{%
 \ifx #1\expandafter \@firstoftwo
 \else \expandafter \@secondoftwo
 \fi
}%
\providecommand \natexlab [1]{#1}%
\providecommand \enquote  [1]{``#1''}%
\providecommand \bibnamefont  [1]{#1}%
\providecommand \bibfnamefont [1]{#1}%
\providecommand \citenamefont [1]{#1}%
\providecommand \href@noop [0]{\@secondoftwo}%
\providecommand \href [0]{\begingroup \@sanitize@url \@href}%
\providecommand \@href[1]{\@@startlink{#1}\@@href}%
\providecommand \@@href[1]{\endgroup#1\@@endlink}%
\providecommand \@sanitize@url [0]{\catcode `\\12\catcode `\$12\catcode
  `\&12\catcode `\#12\catcode `\^12\catcode `\_12\catcode `\%12\relax}%
\providecommand \@@startlink[1]{}%
\providecommand \@@endlink[0]{}%
\providecommand \url  [0]{\begingroup\@sanitize@url \@url }%
\providecommand \@url [1]{\endgroup\@href {#1}{\urlprefix }}%
\providecommand \urlprefix  [0]{URL }%
\providecommand \Eprint [0]{\href }%
\providecommand \doibase [0]{https://doi.org/}%
\providecommand \selectlanguage [0]{\@gobble}%
\providecommand \bibinfo  [0]{\@secondoftwo}%
\providecommand \bibfield  [0]{\@secondoftwo}%
\providecommand \translation [1]{[#1]}%
\providecommand \BibitemOpen [0]{}%
\providecommand \bibitemStop [0]{}%
\providecommand \bibitemNoStop [0]{.\EOS\space}%
\providecommand \EOS [0]{\spacefactor3000\relax}%
\providecommand \BibitemShut  [1]{\csname bibitem#1\endcsname}%
\let\auto@bib@innerbib\@empty
\bibitem [{\citenamefont {Cao}\ \emph {et~al.}(2019)\citenamefont {Cao},
  \citenamefont {Romero}, \citenamefont {Olson}, \citenamefont {Degroote},
  \citenamefont {Johnson}, \citenamefont {Kieferov{\'a}}, \citenamefont
  {Kivlichan}, \citenamefont {Menke}, \citenamefont {Peropadre}, \citenamefont
  {Sawaya} \emph {et~al.}}]{cao2019quantum}%
  \BibitemOpen
  \bibfield  {author} {\bibinfo {author} {\bibfnamefont {Y.}~\bibnamefont
  {Cao}}, \bibinfo {author} {\bibfnamefont {J.}~\bibnamefont {Romero}},
  \bibinfo {author} {\bibfnamefont {J.~P.}\ \bibnamefont {Olson}}, \bibinfo
  {author} {\bibfnamefont {M.}~\bibnamefont {Degroote}}, \bibinfo {author}
  {\bibfnamefont {P.~D.}\ \bibnamefont {Johnson}}, \bibinfo {author}
  {\bibfnamefont {M.}~\bibnamefont {Kieferov{\'a}}}, \bibinfo {author}
  {\bibfnamefont {I.~D.}\ \bibnamefont {Kivlichan}}, \bibinfo {author}
  {\bibfnamefont {T.}~\bibnamefont {Menke}}, \bibinfo {author} {\bibfnamefont
  {B.}~\bibnamefont {Peropadre}}, \bibinfo {author} {\bibfnamefont {N.~P.~D.}\
  \bibnamefont {Sawaya}}, \emph {et~al.},\ }\bibfield  {title} {\bibinfo
  {title} {{Quantum chemistry in the age of quantum computing}},\ }\href
  {https://pubs.acs.org/doi/10.1021/acs.chemrev.8b00803} {\bibfield  {journal}
  {\bibinfo  {journal} {Chem. Rev.}\ }\textbf {\bibinfo {volume} {119}},\
  \bibinfo {pages} {10856} (\bibinfo {year} {2019})}\BibitemShut {NoStop}%
\bibitem [{\citenamefont {Hermann}\ \emph {et~al.}(2017)\citenamefont
  {Hermann}, \citenamefont {DiStasio},\ and\ \citenamefont
  {Tkatchenko}}]{hermann2017first}%
  \BibitemOpen
  \bibfield  {author} {\bibinfo {author} {\bibfnamefont {J.}~\bibnamefont
  {Hermann}}, \bibinfo {author} {\bibfnamefont {R.~A.}\ \bibnamefont
  {DiStasio}},\ and\ \bibinfo {author} {\bibfnamefont {A.}~\bibnamefont
  {Tkatchenko}},\ }\bibfield  {title} {\bibinfo {title} {First-principles
  models for van der waals interactions in molecules and materials: Concepts,
  theory, and applications},\ }\href
  {https://pubs.acs.org/doi/10.1021/acs.chemrev.6b00446} {\bibfield  {journal}
  {\bibinfo  {journal} {Chem. Rev.}\ }\textbf {\bibinfo {volume} {117}},\
  \bibinfo {pages} {4714} (\bibinfo {year} {2017})}\BibitemShut {NoStop}%
\bibitem [{\citenamefont {St{\"o}hr}\ \emph {et~al.}(2019)\citenamefont
  {St{\"o}hr}, \citenamefont {Voorhis},\ and\ \citenamefont
  {Tkatchenko}}]{stohr2019theory}%
  \BibitemOpen
  \bibfield  {author} {\bibinfo {author} {\bibfnamefont {M.}~\bibnamefont
  {St{\"o}hr}}, \bibinfo {author} {\bibfnamefont {T.~V.}\ \bibnamefont
  {Voorhis}},\ and\ \bibinfo {author} {\bibfnamefont {A.}~\bibnamefont
  {Tkatchenko}},\ }\bibfield  {title} {\bibinfo {title} {Theory and practice of
  modeling van der waals interactions in electronic-structure calculations},\
  }\href {https://pubs.rsc.org/en/content/articlelanding/2019/CS/C9CS00060G}
  {\bibfield  {journal} {\bibinfo  {journal} {Chem. Soc. Rev.}\ }\textbf
  {\bibinfo {volume} {48}},\ \bibinfo {pages} {4118} (\bibinfo {year}
  {2019})}\BibitemShut {NoStop}%
\bibitem [{\citenamefont {Margenau}\ and\ \citenamefont
  {Kestner}(2013)}]{margenau2013theory}%
  \BibitemOpen
  \bibfield  {author} {\bibinfo {author} {\bibfnamefont {H.}~\bibnamefont
  {Margenau}}\ and\ \bibinfo {author} {\bibfnamefont {N.~R.}\ \bibnamefont
  {Kestner}},\ }\href@noop {} {\emph {\bibinfo {title} {Theory of
  Intermolecular Forces: International Series of Monographs in Natural
  Philosophy}}},\ edited by\ \bibinfo {editor} {\bibfnamefont {D.~T.}\
  \bibnamefont {Haar}},\ Vol.~\bibinfo {volume} {18}\ (\bibinfo  {publisher}
  {Elsevier},\ \bibinfo {address} {Oxford},\ \bibinfo {year}
  {2013})\BibitemShut {NoStop}%
\bibitem [{\citenamefont {Kaplan}(2006)}]{kaplan2006intermolecular}%
  \BibitemOpen
  \bibfield  {author} {\bibinfo {author} {\bibfnamefont {I.~G.}\ \bibnamefont
  {Kaplan}},\ }\href@noop {} {\emph {\bibinfo {title} {Intermolecular
  interactions: physical picture, computational methods and model
  potentials}}}\ (\bibinfo  {publisher} {John Wiley \& Sons},\ \bibinfo {year}
  {2006})\BibitemShut {NoStop}%
\bibitem [{\citenamefont {Stone}(2013)}]{stone2013theory}%
  \BibitemOpen
  \bibfield  {author} {\bibinfo {author} {\bibfnamefont {A.}~\bibnamefont
  {Stone}},\ }\href@noop {} {\emph {\bibinfo {title} {The theory of
  intermolecular forces}}}\ (\bibinfo  {publisher} {oUP oxford},\ \bibinfo
  {year} {2013})\BibitemShut {NoStop}%
\bibitem [{\citenamefont
  {Hirschfelder}(2009)}]{hirschfelder2009intermolecular}%
  \BibitemOpen
  \bibfield  {author} {\bibinfo {author} {\bibfnamefont {J.~O.}\ \bibnamefont
  {Hirschfelder}},\ }\href@noop {} {\emph {\bibinfo {title} {Intermolecular
  Forces, Volume 12}}},\ edited by\ \bibinfo {editor} {\bibfnamefont
  {I.}~\bibnamefont {Prigogine}}\ and\ \bibinfo {editor} {\bibfnamefont
  {S.}~\bibnamefont {Rice}},\ Vol.~\bibinfo {volume} {12}\ (\bibinfo
  {publisher} {John Wiley \& Sons},\ \bibinfo {year} {2009})\BibitemShut
  {NoStop}%
\bibitem [{\citenamefont {Casimir}\ and\ \citenamefont
  {Polder}(1948)}]{casimir1948influence}%
  \BibitemOpen
  \bibfield  {author} {\bibinfo {author} {\bibfnamefont {H.~B.~G.}\
  \bibnamefont {Casimir}}\ and\ \bibinfo {author} {\bibfnamefont
  {D.}~\bibnamefont {Polder}},\ }\bibfield  {title} {\bibinfo {title} {The
  influence of retardation on the london-van der waals forces},\ }\href
  {https://journals.aps.org/pr/abstract/10.1103/PhysRev.73.360} {\bibfield
  {journal} {\bibinfo  {journal} {Phys. Rev.}\ }\textbf {\bibinfo {volume}
  {73}},\ \bibinfo {pages} {360} (\bibinfo {year} {1948})}\BibitemShut
  {NoStop}%
\bibitem [{\citenamefont {Buhmann}(2013)}]{buhmann2013dispersion}%
  \BibitemOpen
  \bibfield  {author} {\bibinfo {author} {\bibfnamefont {S.~Y.}\ \bibnamefont
  {Buhmann}},\ }\href@noop {} {\emph {\bibinfo {title} {Dispersion Forces I:
  Macroscopic quantum electrodynamics and ground-state Casimir, Casimir--Polder
  and van der Waals forces}}},\ Vol.\ \bibinfo {volume} {247}\ (\bibinfo
  {publisher} {Springer},\ \bibinfo {year} {2013})\BibitemShut {NoStop}%
\bibitem [{\citenamefont {Buhmann}\ and\ \citenamefont
  {Welsch}(2007)}]{buhmann2007dispersion}%
  \BibitemOpen
  \bibfield  {author} {\bibinfo {author} {\bibfnamefont {S.~Y.}\ \bibnamefont
  {Buhmann}}\ and\ \bibinfo {author} {\bibfnamefont {D.~G.}\ \bibnamefont
  {Welsch}},\ }\bibfield  {title} {\bibinfo {title} {Dispersion forces in
  macroscopic quantum electrodynamics},\ }\href
  {https://www.sciencedirect.com/science/article/pii/S0079672707000249}
  {\bibfield  {journal} {\bibinfo  {journal} {Prog. Quantum. Electron.}\
  }\textbf {\bibinfo {volume} {31}},\ \bibinfo {pages} {51} (\bibinfo {year}
  {2007})}\BibitemShut {NoStop}%
\bibitem [{\citenamefont {Compagno}\ \emph {et~al.}(1995)\citenamefont
  {Compagno}, \citenamefont {Passante},\ and\ \citenamefont
  {Persico}}]{compagno1995atom}%
  \BibitemOpen
  \bibfield  {author} {\bibinfo {author} {\bibfnamefont {G.}~\bibnamefont
  {Compagno}}, \bibinfo {author} {\bibfnamefont {R.}~\bibnamefont {Passante}},\
  and\ \bibinfo {author} {\bibfnamefont {F.}~\bibnamefont {Persico}},\
  }\href@noop {} {\emph {\bibinfo {title} {Atom-field interactions and dressed
  atoms}}}\ (\bibinfo  {publisher} {Cambridge University Press},\ \bibinfo
  {address} {New York},\ \bibinfo {year} {1995})\BibitemShut {NoStop}%
\bibitem [{\citenamefont {Passante}(2018)}]{passante2018dispersion}%
  \BibitemOpen
  \bibfield  {author} {\bibinfo {author} {\bibfnamefont {R.}~\bibnamefont
  {Passante}},\ }\bibfield  {title} {\bibinfo {title} {Dispersion interactions
  between neutral atoms and the quantum electrodynamical vacuum},\ }\href
  {https://www.mdpi.com/2073-8994/10/12/735} {\bibfield  {journal} {\bibinfo
  {journal} {Symmetry}\ }\textbf {\bibinfo {volume} {10}},\ \bibinfo {pages}
  {735} (\bibinfo {year} {2018})}\BibitemShut {NoStop}%
\bibitem [{\citenamefont {Cohen-Tannoudji}\ \emph {et~al.}(1997)\citenamefont
  {Cohen-Tannoudji}, \citenamefont {Dupont-Roc},\ and\ \citenamefont
  {Grynberg}}]{cohen1997photons}%
  \BibitemOpen
  \bibfield  {author} {\bibinfo {author} {\bibfnamefont {C.}~\bibnamefont
  {Cohen-Tannoudji}}, \bibinfo {author} {\bibfnamefont {J.}~\bibnamefont
  {Dupont-Roc}},\ and\ \bibinfo {author} {\bibfnamefont {G.}~\bibnamefont
  {Grynberg}},\ }\href@noop {} {\emph {\bibinfo {title} {Photons and
  Atoms-Introduction to Quantum Electrodynamics}}}\ (\bibinfo  {publisher}
  {WILEY‐VCH Verlag GmbH \& Co. KGaA},\ \bibinfo {address} {Morlenbach},\
  \bibinfo {year} {1997})\BibitemShut {NoStop}%
\bibitem [{\citenamefont {Cohen-Tannoudji}\ \emph {et~al.}(1998)\citenamefont
  {Cohen-Tannoudji}, \citenamefont {Dupont-Roc},\ and\ \citenamefont
  {Grynberg}}]{cohen1998atom}%
  \BibitemOpen
  \bibfield  {author} {\bibinfo {author} {\bibfnamefont {C.}~\bibnamefont
  {Cohen-Tannoudji}}, \bibinfo {author} {\bibfnamefont {J.}~\bibnamefont
  {Dupont-Roc}},\ and\ \bibinfo {author} {\bibfnamefont {G.}~\bibnamefont
  {Grynberg}},\ }\href@noop {} {\emph {\bibinfo {title} {Atom-photon
  interactions: basic processes and applications}}}\ (\bibinfo  {publisher}
  {WILEY‐VCH Verlag GmbH \& Co. KGaA},\ \bibinfo {address} {Morlenbach},\
  \bibinfo {year} {1998})\BibitemShut {NoStop}%
\bibitem [{\citenamefont {Preparata}(1995)}]{bookpreparata}%
  \BibitemOpen
  \bibfield  {author} {\bibinfo {author} {\bibfnamefont {G.}~\bibnamefont
  {Preparata}},\ }\href@noop {} {\emph {\bibinfo {title} {{QED Coherence in
  matter}}}}\ (\bibinfo  {publisher} {World Scientific},\ \bibinfo {address}
  {Singapore},\ \bibinfo {year} {1995})\BibitemShut {NoStop}%
\bibitem [{\citenamefont {Salam}(2009)}]{salam2009molecular}%
  \BibitemOpen
  \bibfield  {author} {\bibinfo {author} {\bibfnamefont {A.}~\bibnamefont
  {Salam}},\ }\href@noop {} {\emph {\bibinfo {title} {Molecular quantum
  electrodynamics: long-range intermolecular interactions}}}\ (\bibinfo
  {publisher} {John Wiley \& Sons},\ \bibinfo {year} {2009})\BibitemShut
  {NoStop}%
\bibitem [{\citenamefont {Craig}\ and\ \citenamefont
  {Thirunamachandran}(1984)}]{craig1998t}%
  \BibitemOpen
  \bibfield  {author} {\bibinfo {author} {\bibfnamefont {D.~P.}\ \bibnamefont
  {Craig}}\ and\ \bibinfo {author} {\bibfnamefont {T.}~\bibnamefont
  {Thirunamachandran}},\ }\href@noop {} {\emph {\bibinfo {title} {Molecular
  Quantum Electrodynamics}}}\ (\bibinfo  {publisher} {Academic Press},\
  \bibinfo {address} {London},\ \bibinfo {year} {1984})\BibitemShut {NoStop}%
\bibitem [{\citenamefont {Hoja}\ \emph {et~al.}(2019)\citenamefont {Hoja},
  \citenamefont {Ko}, \citenamefont {Neumann}, \citenamefont {Car},
  \citenamefont {DiStasio},\ and\ \citenamefont
  {Tkatchenko}}]{hoja2019reliable}%
  \BibitemOpen
  \bibfield  {author} {\bibinfo {author} {\bibfnamefont {J.}~\bibnamefont
  {Hoja}}, \bibinfo {author} {\bibfnamefont {H.-Y.}\ \bibnamefont {Ko}},
  \bibinfo {author} {\bibfnamefont {M.~A.}\ \bibnamefont {Neumann}}, \bibinfo
  {author} {\bibfnamefont {R.}~\bibnamefont {Car}}, \bibinfo {author}
  {\bibfnamefont {R.~A.}\ \bibnamefont {DiStasio}},\ and\ \bibinfo {author}
  {\bibfnamefont {A.}~\bibnamefont {Tkatchenko}},\ }\bibfield  {title}
  {\bibinfo {title} {Reliable and practical computational description of
  molecular crystal polymorphs},\ }\href
  {https://doi.org/10.1126/sciadv.aau3338} {\bibfield  {journal} {\bibinfo
  {journal} {Sci. Adv.}\ }\textbf {\bibinfo {volume} {5}},\ \bibinfo {pages}
  {eaau3338} (\bibinfo {year} {2019})}\BibitemShut {NoStop}%
\bibitem [{\citenamefont {Hoja}\ and\ \citenamefont
  {Tkatchenko}(2018)}]{hoja2018first}%
  \BibitemOpen
  \bibfield  {author} {\bibinfo {author} {\bibfnamefont {J.}~\bibnamefont
  {Hoja}}\ and\ \bibinfo {author} {\bibfnamefont {A.}~\bibnamefont
  {Tkatchenko}},\ }\bibfield  {title} {\bibinfo {title} {First-principles
  stability ranking of molecular crystal polymorphs with the dft+ mbd
  approach},\ }\href
  {https://pubs.rsc.org/en/content/articlelanding/2018/fd/c8fd00066b}
  {\bibfield  {journal} {\bibinfo  {journal} {Faraday Discuss.}\ }\textbf
  {\bibinfo {volume} {211}},\ \bibinfo {pages} {253} (\bibinfo {year}
  {2018})}\BibitemShut {NoStop}%
\bibitem [{\citenamefont {Mortazavi}\ \emph {et~al.}(2018)\citenamefont
  {Mortazavi}, \citenamefont {Brandenburg}, \citenamefont {Maurer},\ and\
  \citenamefont {Tkatchenko}}]{mortazavi2018structure}%
  \BibitemOpen
  \bibfield  {author} {\bibinfo {author} {\bibfnamefont {M.}~\bibnamefont
  {Mortazavi}}, \bibinfo {author} {\bibfnamefont {J.~G.}\ \bibnamefont
  {Brandenburg}}, \bibinfo {author} {\bibfnamefont {R.~J.}\ \bibnamefont
  {Maurer}},\ and\ \bibinfo {author} {\bibfnamefont {A.}~\bibnamefont
  {Tkatchenko}},\ }\bibfield  {title} {\bibinfo {title} {Structure and
  stability of molecular crystals with many-body dispersion-inclusive density
  functional tight binding},\ }\href
  {https://pubs.acs.org/doi/abs/10.1021/acs.jpclett.7b03234} {\bibfield
  {journal} {\bibinfo  {journal} {J. Phys. Chem. Lett.}\ }\textbf {\bibinfo
  {volume} {9}},\ \bibinfo {pages} {399} (\bibinfo {year} {2018})}\BibitemShut
  {NoStop}%
\bibitem [{\citenamefont {St{\"o}hr}\ and\ \citenamefont
  {Tkatchenko}(2019)}]{stohr2019quantum}%
  \BibitemOpen
  \bibfield  {author} {\bibinfo {author} {\bibfnamefont {M.}~\bibnamefont
  {St{\"o}hr}}\ and\ \bibinfo {author} {\bibfnamefont {A.}~\bibnamefont
  {Tkatchenko}},\ }\bibfield  {title} {\bibinfo {title} {Quantum mechanics of
  proteins in explicit water: The role of plasmon-like solute-solvent
  interactions},\ }\href {https://www.science.org/doi/10.1126/sciadv.aax0024}
  {\bibfield  {journal} {\bibinfo  {journal} {Sci. Adv.}\ }\textbf {\bibinfo
  {volume} {5}},\ \bibinfo {pages} {eaax0024} (\bibinfo {year}
  {2019})}\BibitemShut {NoStop}%
\bibitem [{\citenamefont {Reilly}\ and\ \citenamefont
  {Tkatchenko}(2014)}]{reilly2014role}%
  \BibitemOpen
  \bibfield  {author} {\bibinfo {author} {\bibfnamefont {A.~M.}\ \bibnamefont
  {Reilly}}\ and\ \bibinfo {author} {\bibfnamefont {A.}~\bibnamefont
  {Tkatchenko}},\ }\bibfield  {title} {\bibinfo {title} {Role of dispersion
  interactions in the polymorphism and entropic stabilization of the aspirin
  crystal},\ }\href
  {https://journals.aps.org/prl/abstract/10.1103/PhysRevLett.113.055701}
  {\bibfield  {journal} {\bibinfo  {journal} {Phys. Rev. Lett.}\ }\textbf
  {\bibinfo {volume} {113}},\ \bibinfo {pages} {055701} (\bibinfo {year}
  {2014})}\BibitemShut {NoStop}%
\bibitem [{\citenamefont {Galante}\ and\ \citenamefont
  {Tkatchenko}(2023)}]{PhysRevResearch.5.L012028}%
  \BibitemOpen
  \bibfield  {author} {\bibinfo {author} {\bibfnamefont {M.}~\bibnamefont
  {Galante}}\ and\ \bibinfo {author} {\bibfnamefont {A.}~\bibnamefont
  {Tkatchenko}},\ }\bibfield  {title} {\bibinfo {title} {Anisotropic van der
  waals dispersion forces in polymers: Structural symmetry breaking leads to
  enhanced conformational search},\ }\href
  {https://doi.org/10.1103/PhysRevResearch.5.L012028} {\bibfield  {journal}
  {\bibinfo  {journal} {Phys. Rev. Res.}\ }\textbf {\bibinfo {volume} {5}},\
  \bibinfo {pages} {L012028} (\bibinfo {year} {2023})}\BibitemShut {NoStop}%
\bibitem [{\citenamefont {Kleshchonok}\ and\ \citenamefont
  {Tkatchenko}(2018)}]{kleshchonok2018tailoring}%
  \BibitemOpen
  \bibfield  {author} {\bibinfo {author} {\bibfnamefont {A.}~\bibnamefont
  {Kleshchonok}}\ and\ \bibinfo {author} {\bibfnamefont {A.}~\bibnamefont
  {Tkatchenko}},\ }\bibfield  {title} {\bibinfo {title} {Tailoring van der
  waals dispersion interactions with external electric charges},\ }\href
  {https://www.nature.com/articles/s41467-018-05407-x} {\bibfield  {journal}
  {\bibinfo  {journal} {Nat. Commun.}\ }\textbf {\bibinfo {volume} {9}},\
  \bibinfo {pages} {3017} (\bibinfo {year} {2018})}\BibitemShut {NoStop}%
\bibitem [{\citenamefont {Ambrosetti}\ \emph {et~al.}(2022)\citenamefont
  {Ambrosetti}, \citenamefont {Umari}, \citenamefont {Silvestrelli},
  \citenamefont {Elliott},\ and\ \citenamefont
  {Tkatchenko}}]{ambrosetti2022optical}%
  \BibitemOpen
  \bibfield  {author} {\bibinfo {author} {\bibfnamefont {A.}~\bibnamefont
  {Ambrosetti}}, \bibinfo {author} {\bibfnamefont {P.}~\bibnamefont {Umari}},
  \bibinfo {author} {\bibfnamefont {P.~L.}\ \bibnamefont {Silvestrelli}},
  \bibinfo {author} {\bibfnamefont {J.}~\bibnamefont {Elliott}},\ and\ \bibinfo
  {author} {\bibfnamefont {A.}~\bibnamefont {Tkatchenko}},\ }\bibfield  {title}
  {\bibinfo {title} {Optical van-der-waals forces in molecules: from electronic
  bethe-salpeter calculations to the many-body dispersion model},\ }\href
  {https://www.nature.com/articles/s41467-022-28461-y} {\bibfield  {journal}
  {\bibinfo  {journal} {Nat. Commun.}\ }\textbf {\bibinfo {volume} {13}},\
  \bibinfo {pages} {813} (\bibinfo {year} {2022})}\BibitemShut {NoStop}%
\bibitem [{\citenamefont {Karimpour}\ \emph {et~al.}(2022)\citenamefont
  {Karimpour}, \citenamefont {Fedorov},\ and\ \citenamefont
  {Tkatchenko}}]{Karimpour_2022}%
  \BibitemOpen
  \bibfield  {author} {\bibinfo {author} {\bibfnamefont {M.~R.}\ \bibnamefont
  {Karimpour}}, \bibinfo {author} {\bibfnamefont {D.~V.}\ \bibnamefont
  {Fedorov}},\ and\ \bibinfo {author} {\bibfnamefont {A.}~\bibnamefont
  {Tkatchenko}},\ }\bibfield  {title} {\bibinfo {title} {Quantum framework for
  describing retarded and nonretarded molecular interactions in external
  electric fields},\ }\href {https://doi.org/10.1103/physrevresearch.4.013011}
  {\bibfield  {journal} {\bibinfo  {journal} {Phys. Rev. Res.}\ }\textbf
  {\bibinfo {volume} {4}},\ \bibinfo {pages} {013011} (\bibinfo {year}
  {2022})}\BibitemShut {NoStop}%
\bibitem [{\citenamefont {Richardson}(1975)}]{richardson1975dispersion}%
  \BibitemOpen
  \bibfield  {author} {\bibinfo {author} {\bibfnamefont {D.~D.}\ \bibnamefont
  {Richardson}},\ }\bibfield  {title} {\bibinfo {title} {Dispersion
  contribution of two-atom interaction energy: Multipole interactions},\ }\href
  {https://iopscience.iop.org/article/10.1088/0305-4470/8/11/019} {\bibfield
  {journal} {\bibinfo  {journal} {J. Phys. A: Math}\ }\textbf {\bibinfo
  {volume} {8}},\ \bibinfo {pages} {1828} (\bibinfo {year} {1975})}\BibitemShut
  {NoStop}%
\bibitem [{\citenamefont {Mahanty}\ and\ \citenamefont
  {Ninham}(1973)}]{mahanty1973dispersion}%
  \BibitemOpen
  \bibfield  {author} {\bibinfo {author} {\bibfnamefont {J.}~\bibnamefont
  {Mahanty}}\ and\ \bibinfo {author} {\bibfnamefont {B.~W.}\ \bibnamefont
  {Ninham}},\ }\bibfield  {title} {\bibinfo {title} {Dispersion contributions
  to surface energy},\ }\href
  {https://pubs.aip.org/aip/jcp/article-abstract/59/11/6157/213316/Dispersion-contributions-to-surface-energy?redirectedFrom=fulltext}
  {\bibfield  {journal} {\bibinfo  {journal} {J. Chem. Phys.}\ }\textbf
  {\bibinfo {volume} {59}},\ \bibinfo {pages} {6157} (\bibinfo {year}
  {1973})}\BibitemShut {NoStop}%
\bibitem [{\citenamefont {Woods}\ \emph {et~al.}(2016)\citenamefont {Woods},
  \citenamefont {Dalvit}, \citenamefont {Tkatchenko}, \citenamefont
  {Rodriguez-Lopez}, \citenamefont {Rodriguez},\ and\ \citenamefont
  {Podgornik}}]{RevModPhys.88.045003}%
  \BibitemOpen
  \bibfield  {author} {\bibinfo {author} {\bibfnamefont {L.~M.}\ \bibnamefont
  {Woods}}, \bibinfo {author} {\bibfnamefont {D.~A.~R.}\ \bibnamefont
  {Dalvit}}, \bibinfo {author} {\bibfnamefont {A.}~\bibnamefont {Tkatchenko}},
  \bibinfo {author} {\bibfnamefont {P.}~\bibnamefont {Rodriguez-Lopez}},
  \bibinfo {author} {\bibfnamefont {A.~W.}\ \bibnamefont {Rodriguez}},\ and\
  \bibinfo {author} {\bibfnamefont {R.}~\bibnamefont {Podgornik}},\ }\bibfield
  {title} {\bibinfo {title} {Materials perspective on casimir and van der waals
  interactions},\ }\href {https://doi.org/10.1103/RevModPhys.88.045003}
  {\bibfield  {journal} {\bibinfo  {journal} {Rev. Mod. Phys.}\ }\textbf
  {\bibinfo {volume} {88}},\ \bibinfo {pages} {045003} (\bibinfo {year}
  {2016})}\BibitemShut {NoStop}%
\bibitem [{\citenamefont {Tkatchenko}(2015)}]{tkatchenko2015current}%
  \BibitemOpen
  \bibfield  {author} {\bibinfo {author} {\bibfnamefont {A.}~\bibnamefont
  {Tkatchenko}},\ }\bibfield  {title} {\bibinfo {title} {Current understanding
  of van der waals effects in realistic materials},\ }\href
  {https://onlinelibrary.wiley.com/doi/10.1002/adfm.201403029} {\bibfield
  {journal} {\bibinfo  {journal} {Adv. Funct. Mater.}\ }\textbf {\bibinfo
  {volume} {25}},\ \bibinfo {pages} {2054} (\bibinfo {year}
  {2015})}\BibitemShut {NoStop}%
\bibitem [{\citenamefont {Ren}\ \emph {et~al.}(2012)\citenamefont {Ren},
  \citenamefont {Rinke}, \citenamefont {Joas},\ and\ \citenamefont
  {Scheffler}}]{ren2012random}%
  \BibitemOpen
  \bibfield  {author} {\bibinfo {author} {\bibfnamefont {X.}~\bibnamefont
  {Ren}}, \bibinfo {author} {\bibfnamefont {P.}~\bibnamefont {Rinke}}, \bibinfo
  {author} {\bibfnamefont {C.}~\bibnamefont {Joas}},\ and\ \bibinfo {author}
  {\bibfnamefont {M.}~\bibnamefont {Scheffler}},\ }\bibfield  {title} {\bibinfo
  {title} {Random-phase approximation and its applications in computational
  chemistry and materials science},\ }\href
  {https://link.springer.com/article/10.1007/s10853-012-6570-4} {\bibfield
  {journal} {\bibinfo  {journal} {J. Mater. Sci.}\ }\textbf {\bibinfo {volume}
  {47}},\ \bibinfo {pages} {7447} (\bibinfo {year} {2012})}\BibitemShut
  {NoStop}%
\bibitem [{\citenamefont {Harl}\ and\ \citenamefont
  {Kresse}(2009)}]{harl2009accurate}%
  \BibitemOpen
  \bibfield  {author} {\bibinfo {author} {\bibfnamefont {J.}~\bibnamefont
  {Harl}}\ and\ \bibinfo {author} {\bibfnamefont {G.}~\bibnamefont {Kresse}},\
  }\bibfield  {title} {\bibinfo {title} {Accurate bulk properties from
  approximate many-body techniques},\ }\href
  {https://pubmed.ncbi.nlm.nih.gov/19792517/} {\bibfield  {journal} {\bibinfo
  {journal} {Phys. Rev. Lett.}\ }\textbf {\bibinfo {volume} {103}},\ \bibinfo
  {pages} {056401} (\bibinfo {year} {2009})}\BibitemShut {NoStop}%
\bibitem [{\citenamefont {Dobson}\ and\ \citenamefont
  {Gould}(2012)}]{dobson2012calculation}%
  \BibitemOpen
  \bibfield  {author} {\bibinfo {author} {\bibfnamefont {J.~F.}\ \bibnamefont
  {Dobson}}\ and\ \bibinfo {author} {\bibfnamefont {T.}~\bibnamefont {Gould}},\
  }\bibfield  {title} {\bibinfo {title} {Calculation of dispersion energies},\
  }\href {https://iopscience.iop.org/article/10.1088/0953-8984/24/7/073201}
  {\bibfield  {journal} {\bibinfo  {journal} {J. Condens. Matter Phys.}\
  }\textbf {\bibinfo {volume} {24}},\ \bibinfo {pages} {073201} (\bibinfo
  {year} {2012})}\BibitemShut {NoStop}%
\bibitem [{\citenamefont {Parsegian}(2005)}]{parsegian2005van}%
  \BibitemOpen
  \bibfield  {author} {\bibinfo {author} {\bibfnamefont {V.~A.}\ \bibnamefont
  {Parsegian}},\ }\href@noop {} {\emph {\bibinfo {title} {Van der Waals forces:
  a handbook for biologists, chemists, engineers, and physicists}}}\ (\bibinfo
  {publisher} {Cambridge university press},\ \bibinfo {year}
  {2005})\BibitemShut {NoStop}%
\bibitem [{\citenamefont {Becke}\ and\ \citenamefont
  {Johnson}(2006{\natexlab{a}})}]{becke2006simple}%
  \BibitemOpen
  \bibfield  {author} {\bibinfo {author} {\bibfnamefont {A.~D.}\ \bibnamefont
  {Becke}}\ and\ \bibinfo {author} {\bibfnamefont {E.~R.}\ \bibnamefont
  {Johnson}},\ }\bibfield  {title} {\bibinfo {title} {A simple effective
  potential for exchange},\ }\href
  {https://pubs.aip.org/aip/jcp/article/124/22/221101/920551/A-simple-effective-potential-for-exchange}
  {\bibfield  {journal} {\bibinfo  {journal} {J. Chem. Phys.}\ }\textbf
  {\bibinfo {volume} {124}},\ \bibinfo {pages} {221101} (\bibinfo {year}
  {2006}{\natexlab{a}})}\BibitemShut {NoStop}%
\bibitem [{\citenamefont {Becke}\ and\ \citenamefont
  {Johnson}(2006{\natexlab{b}})}]{becke2006exchange}%
  \BibitemOpen
  \bibfield  {author} {\bibinfo {author} {\bibfnamefont {A.~D.}\ \bibnamefont
  {Becke}}\ and\ \bibinfo {author} {\bibfnamefont {E.~R.}\ \bibnamefont
  {Johnson}},\ }\bibfield  {title} {\bibinfo {title} {Exchange-hole dipole
  moment and the dispersion interaction: High-order dispersion coefficients},\
  }\href
  {https://pubs.aip.org/aip/jcp/article-abstract/124/1/014104/897786/Exchange-hole-dipole-moment-and-the-dispersion?redirectedFrom=fulltext}
  {\bibfield  {journal} {\bibinfo  {journal} {J. Chem. Phys.}\ }\textbf
  {\bibinfo {volume} {124}},\ \bibinfo {pages} {014104} (\bibinfo {year}
  {2006}{\natexlab{b}})}\BibitemShut {NoStop}%
\bibitem [{\citenamefont {Grimme}\ \emph {et~al.}(2010)\citenamefont {Grimme},
  \citenamefont {Antony}, \citenamefont {Ehrlich},\ and\ \citenamefont
  {Krieg}}]{grimme2010consistent}%
  \BibitemOpen
  \bibfield  {author} {\bibinfo {author} {\bibfnamefont {S.}~\bibnamefont
  {Grimme}}, \bibinfo {author} {\bibfnamefont {J.}~\bibnamefont {Antony}},
  \bibinfo {author} {\bibfnamefont {S.}~\bibnamefont {Ehrlich}},\ and\ \bibinfo
  {author} {\bibfnamefont {H.}~\bibnamefont {Krieg}},\ }\bibfield  {title}
  {\bibinfo {title} {A consistent and accurate ab initio parametrization of
  density functional dispersion correction (dft-d) for the 94 elements h-pu},\
  }\href
  {https://pubs.aip.org/aip/jcp/article/132/15/154104/926936/A-consistent-and-accurate-ab-initio}
  {\bibfield  {journal} {\bibinfo  {journal} {J. Chem. Phys.}\ }\textbf
  {\bibinfo {volume} {132}},\ \bibinfo {pages} {154104} (\bibinfo {year}
  {2010})}\BibitemShut {NoStop}%
\bibitem [{\citenamefont {Grimme}(2006)}]{grimme2006semiempirical}%
  \BibitemOpen
  \bibfield  {author} {\bibinfo {author} {\bibfnamefont {S.}~\bibnamefont
  {Grimme}},\ }\bibfield  {title} {\bibinfo {title} {Semiempirical gga-type
  density functional constructed with a long-range dispersion correction},\
  }\href {https://onlinelibrary.wiley.com/doi/abs/10.1002/jcc.20495} {\bibfield
   {journal} {\bibinfo  {journal} {J. Comput. Chem.}\ }\textbf {\bibinfo
  {volume} {27}},\ \bibinfo {pages} {1787} (\bibinfo {year}
  {2006})}\BibitemShut {NoStop}%
\bibitem [{\citenamefont {Tkatchenko}\ \emph {et~al.}(2012)\citenamefont
  {Tkatchenko}, \citenamefont {DiStasio}, \citenamefont {Car},\ and\
  \citenamefont {Scheffler}}]{tkatchenko2012accurate}%
  \BibitemOpen
  \bibfield  {author} {\bibinfo {author} {\bibfnamefont {A.}~\bibnamefont
  {Tkatchenko}}, \bibinfo {author} {\bibfnamefont {R.~A.}\ \bibnamefont
  {DiStasio}}, \bibinfo {author} {\bibfnamefont {R.}~\bibnamefont {Car}},\ and\
  \bibinfo {author} {\bibfnamefont {M.}~\bibnamefont {Scheffler}},\ }\bibfield
  {title} {\bibinfo {title} {Accurate and efficient method for many-body van
  der waals interactions},\ }\href
  {https://journals.aps.org/prl/abstract/10.1103/PhysRevLett.108.236402}
  {\bibfield  {journal} {\bibinfo  {journal} {Phys. Rev. Lett.}\ }\textbf
  {\bibinfo {volume} {108}},\ \bibinfo {pages} {236402} (\bibinfo {year}
  {2012})}\BibitemShut {NoStop}%
\bibitem [{\citenamefont {Massa}\ \emph
  {et~al.}(2021{\natexlab{a}})\citenamefont {Massa}, \citenamefont
  {Ambrosetti},\ and\ \citenamefont {Silvestrelli}}]{massa2021many}%
  \BibitemOpen
  \bibfield  {author} {\bibinfo {author} {\bibfnamefont {D.}~\bibnamefont
  {Massa}}, \bibinfo {author} {\bibfnamefont {A.}~\bibnamefont {Ambrosetti}},\
  and\ \bibinfo {author} {\bibfnamefont {P.~L.}\ \bibnamefont {Silvestrelli}},\
  }\bibfield  {title} {\bibinfo {title} {Many-body van der waals interactions
  beyond the dipole approximation},\ }\href
  {https://pubs.aip.org/aip/jcp/article/154/22/224115/313308/Many-body-van-der-Waals-interactions-beyond-the}
  {\bibfield  {journal} {\bibinfo  {journal} {J. Chem. Phys.}\ }\textbf
  {\bibinfo {volume} {154}},\ \bibinfo {pages} {224115} (\bibinfo {year}
  {2021}{\natexlab{a}})}\BibitemShut {NoStop}%
\bibitem [{\citenamefont {Ambrosetti}\ \emph {et~al.}(2014)\citenamefont
  {Ambrosetti}, \citenamefont {Reilly}, \citenamefont {DiStasio},\ and\
  \citenamefont {Tkatchenko}}]{ambrosetti2014long}%
  \BibitemOpen
  \bibfield  {author} {\bibinfo {author} {\bibfnamefont {A.}~\bibnamefont
  {Ambrosetti}}, \bibinfo {author} {\bibfnamefont {A.~M.}\ \bibnamefont
  {Reilly}}, \bibinfo {author} {\bibfnamefont {R.~A.}\ \bibnamefont
  {DiStasio}},\ and\ \bibinfo {author} {\bibfnamefont {A.}~\bibnamefont
  {Tkatchenko}},\ }\bibfield  {title} {\bibinfo {title} {Long-range correlation
  energy calculated from coupled atomic response functions},\ }\href
  {https://pubs.aip.org/aip/jcp/article/140/18/18A508/149387/Long-range-correlation-energy-calculated-from}
  {\bibfield  {journal} {\bibinfo  {journal} {J. Chem. Phys.}\ }\textbf
  {\bibinfo {volume} {140}},\ \bibinfo {pages} {18A508} (\bibinfo {year}
  {2014})}\BibitemShut {NoStop}%
\bibitem [{\citenamefont {Bade}\ and\ \citenamefont
  {Kirkwood}(1957)}]{doi:10.1063/1.1743992}%
  \BibitemOpen
  \bibfield  {author} {\bibinfo {author} {\bibfnamefont {W.~L.}\ \bibnamefont
  {Bade}}\ and\ \bibinfo {author} {\bibfnamefont {J.~G.}\ \bibnamefont
  {Kirkwood}},\ }\bibfield  {title} {\bibinfo {title} {Drude‐model
  calculation of dispersion forces. ii. the linear lattice},\ }\href
  {https://doi.org/10.1063/1.1743992} {\bibfield  {journal} {\bibinfo
  {journal} {J. Chem. Phys.}\ }\textbf {\bibinfo {volume} {27}},\ \bibinfo
  {pages} {1284} (\bibinfo {year} {1957})}\BibitemShut {NoStop}%
\bibitem [{\citenamefont {Gori}\ \emph {et~al.}(2022)\citenamefont {Gori},
  \citenamefont {Kurian},\ and\ \citenamefont
  {Tkatchenko}}]{https://doi.org/10.48550/arxiv.2205.11549}%
  \BibitemOpen
  \bibfield  {author} {\bibinfo {author} {\bibfnamefont {M.}~\bibnamefont
  {Gori}}, \bibinfo {author} {\bibfnamefont {P.}~\bibnamefont {Kurian}},\ and\
  \bibinfo {author} {\bibfnamefont {A.}~\bibnamefont {Tkatchenko}},\ }\bibfield
   {title} {\bibinfo {title} {Second quantization approach to many-body
  dispersion interactions},\ }\href {https://arxiv.org/abs/2205.11549}
  {\bibfield  {journal} {\bibinfo  {journal} {arXiv:2205.11549}\ } (\bibinfo
  {year} {2022})}\BibitemShut {NoStop}%
\bibitem [{\citenamefont {Massa}\ \emph
  {et~al.}(2021{\natexlab{b}})\citenamefont {Massa}, \citenamefont
  {Ambrosetti},\ and\ \citenamefont {Silvestrelli}}]{massa2021beyond}%
  \BibitemOpen
  \bibfield  {author} {\bibinfo {author} {\bibfnamefont {D.}~\bibnamefont
  {Massa}}, \bibinfo {author} {\bibfnamefont {A.}~\bibnamefont {Ambrosetti}},\
  and\ \bibinfo {author} {\bibfnamefont {P.~L.}\ \bibnamefont {Silvestrelli}},\
  }\bibfield  {title} {\bibinfo {title} {Beyond-dipole van der waals
  contributions within the many-body dispersion framework},\ }\href
  {https://iopscience.iop.org/article/10.1088/2516-1075/ac3b5c/meta} {\bibfield
   {journal} {\bibinfo  {journal} {Electron. struct.}\ }\textbf {\bibinfo
  {volume} {3}},\ \bibinfo {pages} {044002} (\bibinfo {year}
  {2021}{\natexlab{b}})}\BibitemShut {NoStop}%
\bibitem [{\citenamefont {Sadhukhan}\ and\ \citenamefont
  {Manby}(2016)}]{sadhukhan2016quantum}%
  \BibitemOpen
  \bibfield  {author} {\bibinfo {author} {\bibfnamefont {M.}~\bibnamefont
  {Sadhukhan}}\ and\ \bibinfo {author} {\bibfnamefont {F.~R.}\ \bibnamefont
  {Manby}},\ }\bibfield  {title} {\bibinfo {title} {Quantum mechanics of drude
  oscillators with full coulomb interaction},\ }\href
  {https://journals.aps.org/prb/abstract/10.1103/PhysRevB.94.115106} {\bibfield
   {journal} {\bibinfo  {journal} {Phys. Rev. B}\ }\textbf {\bibinfo {volume}
  {94}},\ \bibinfo {pages} {115106} (\bibinfo {year} {2016})}\BibitemShut
  {NoStop}%
\bibitem [{\citenamefont {Peruzzo}\ \emph {et~al.}(2014)\citenamefont
  {Peruzzo}, \citenamefont {McClean}, \citenamefont {Shadbolt}, \citenamefont
  {Yung}, \citenamefont {Zhou}, \citenamefont {Love}, \citenamefont
  {Aspuru-Guzik},\ and\ \citenamefont {O’brien}}]{peruzzo2014variational}%
  \BibitemOpen
  \bibfield  {author} {\bibinfo {author} {\bibfnamefont {A.}~\bibnamefont
  {Peruzzo}}, \bibinfo {author} {\bibfnamefont {J.}~\bibnamefont {McClean}},
  \bibinfo {author} {\bibfnamefont {P.}~\bibnamefont {Shadbolt}}, \bibinfo
  {author} {\bibfnamefont {M.-H.}\ \bibnamefont {Yung}}, \bibinfo {author}
  {\bibfnamefont {X.-Q.}\ \bibnamefont {Zhou}}, \bibinfo {author}
  {\bibfnamefont {P.~J.}\ \bibnamefont {Love}}, \bibinfo {author}
  {\bibfnamefont {A.}~\bibnamefont {Aspuru-Guzik}},\ and\ \bibinfo {author}
  {\bibfnamefont {J.~L.}\ \bibnamefont {O’brien}},\ }\bibfield  {title}
  {\bibinfo {title} {A variational eigenvalue solver on a photonic quantum
  processor},\ }\href {https://www.nature.com/articles/ncomms5213} {\bibfield
  {journal} {\bibinfo  {journal} {Nat. Commun.}\ }\textbf {\bibinfo {volume}
  {5}},\ \bibinfo {pages} {4213} (\bibinfo {year} {2014})}\BibitemShut
  {NoStop}%
\bibitem [{\citenamefont {Kandala}\ \emph {et~al.}(2017)\citenamefont
  {Kandala}, \citenamefont {Mezzacapo}, \citenamefont {Temme}, \citenamefont
  {Takita}, \citenamefont {Brink}, \citenamefont {Chow},\ and\ \citenamefont
  {Gambetta}}]{kandala2017hardware}%
  \BibitemOpen
  \bibfield  {author} {\bibinfo {author} {\bibfnamefont {A.}~\bibnamefont
  {Kandala}}, \bibinfo {author} {\bibfnamefont {A.}~\bibnamefont {Mezzacapo}},
  \bibinfo {author} {\bibfnamefont {K.}~\bibnamefont {Temme}}, \bibinfo
  {author} {\bibfnamefont {M.}~\bibnamefont {Takita}}, \bibinfo {author}
  {\bibfnamefont {M.}~\bibnamefont {Brink}}, \bibinfo {author} {\bibfnamefont
  {J.~M.}\ \bibnamefont {Chow}},\ and\ \bibinfo {author} {\bibfnamefont
  {J.~M.}\ \bibnamefont {Gambetta}},\ }\bibfield  {title} {\bibinfo {title}
  {Hardware-efficient variational quantum eigensolver for small molecules and
  quantum magnets},\ }\href {https://www.nature.com/articles/nature23879}
  {\bibfield  {journal} {\bibinfo  {journal} {nature}\ }\textbf {\bibinfo
  {volume} {549}},\ \bibinfo {pages} {242} (\bibinfo {year}
  {2017})}\BibitemShut {NoStop}%
\bibitem [{\citenamefont {Nam}\ \emph {et~al.}(2020)\citenamefont {Nam},
  \citenamefont {Chen}, \citenamefont {Pisenti}, \citenamefont {Wright},
  \citenamefont {Delaney}, \citenamefont {Maslov}, \citenamefont {Brown},
  \citenamefont {Allen}, \citenamefont {Amini} \emph {et~al.}}]{nam2020ground}%
  \BibitemOpen
  \bibfield  {author} {\bibinfo {author} {\bibfnamefont {Y.}~\bibnamefont
  {Nam}}, \bibinfo {author} {\bibfnamefont {J.}~\bibnamefont {Chen}}, \bibinfo
  {author} {\bibfnamefont {N.~C.}\ \bibnamefont {Pisenti}}, \bibinfo {author}
  {\bibfnamefont {K.}~\bibnamefont {Wright}}, \bibinfo {author} {\bibfnamefont
  {C.}~\bibnamefont {Delaney}}, \bibinfo {author} {\bibfnamefont
  {D.}~\bibnamefont {Maslov}}, \bibinfo {author} {\bibfnamefont {K.~R.}\
  \bibnamefont {Brown}}, \bibinfo {author} {\bibfnamefont {S.}~\bibnamefont
  {Allen}}, \bibinfo {author} {\bibfnamefont {J.~M.}\ \bibnamefont {Amini}},
  \emph {et~al.},\ }\bibfield  {title} {\bibinfo {title} {Ground-state energy
  estimation of the water molecule on a trapped-ion quantum computer},\ }\href
  {https://www.nature.com/articles/s41534-020-0259-3} {\bibfield  {journal}
  {\bibinfo  {journal} {Npj Quantum Inf.}\ }\textbf {\bibinfo {volume} {6}},\
  \bibinfo {pages} {33} (\bibinfo {year} {2020})}\BibitemShut {NoStop}%
\bibitem [{\citenamefont {Anderson}\ \emph {et~al.}(2022)\citenamefont
  {Anderson}, \citenamefont {Kiffner}, \citenamefont {Barkoutsos},
  \citenamefont {Tavernelli}, \citenamefont {Crain},\ and\ \citenamefont
  {Jaksch}}]{anderson2022coarse}%
  \BibitemOpen
  \bibfield  {author} {\bibinfo {author} {\bibfnamefont {L.~W.}\ \bibnamefont
  {Anderson}}, \bibinfo {author} {\bibfnamefont {M.}~\bibnamefont {Kiffner}},
  \bibinfo {author} {\bibfnamefont {P.~K.}\ \bibnamefont {Barkoutsos}},
  \bibinfo {author} {\bibfnamefont {I.}~\bibnamefont {Tavernelli}}, \bibinfo
  {author} {\bibfnamefont {J.}~\bibnamefont {Crain}},\ and\ \bibinfo {author}
  {\bibfnamefont {D.}~\bibnamefont {Jaksch}},\ }\bibfield  {title} {\bibinfo
  {title} {Coarse-grained intermolecular interactions on quantum processors},\
  }\href {https://journals.aps.org/pra/abstract/10.1103/PhysRevA.105.062409}
  {\bibfield  {journal} {\bibinfo  {journal} {Phys. Rev. A}\ }\textbf {\bibinfo
  {volume} {105}},\ \bibinfo {pages} {062409} (\bibinfo {year}
  {2022})}\BibitemShut {NoStop}%
\bibitem [{\citenamefont {Ditte}\ \emph {et~al.}(2023)\citenamefont {Ditte},
  \citenamefont {Barborini}, \citenamefont {Sandonas},\ and\ \citenamefont
  {Tkatchenko}}]{ditte2023molecules}%
  \BibitemOpen
  \bibfield  {author} {\bibinfo {author} {\bibfnamefont {M.}~\bibnamefont
  {Ditte}}, \bibinfo {author} {\bibfnamefont {M.}~\bibnamefont {Barborini}},
  \bibinfo {author} {\bibfnamefont {L.~M.}\ \bibnamefont {Sandonas}},\ and\
  \bibinfo {author} {\bibfnamefont {A.}~\bibnamefont {Tkatchenko}},\ }\bibfield
   {title} {\bibinfo {title} {Molecules in environments: Towards systematic
  quantum embedding of electrons and drude oscillators},\ }\href
  {https://arxiv.org/abs/2303.16250} {\bibfield  {journal} {\bibinfo  {journal}
  {arXiv:2303.16250}\ } (\bibinfo {year} {2023})}\BibitemShut {NoStop}%
\bibitem [{\citenamefont {Lloyd}\ and\ \citenamefont
  {Braunstein}(1999)}]{lloyd1999quantum}%
  \BibitemOpen
  \bibfield  {author} {\bibinfo {author} {\bibfnamefont {S.}~\bibnamefont
  {Lloyd}}\ and\ \bibinfo {author} {\bibfnamefont {S.~L.}\ \bibnamefont
  {Braunstein}},\ }\bibfield  {title} {\bibinfo {title} {Quantum computation
  over continuous variables},\ }\href
  {https://journals.aps.org/prl/abstract/10.1103/PhysRevLett.82.1784}
  {\bibfield  {journal} {\bibinfo  {journal} {Phys. Rev. Lett.}\ }\textbf
  {\bibinfo {volume} {82}},\ \bibinfo {pages} {1784} (\bibinfo {year}
  {1999})}\BibitemShut {NoStop}%
\bibitem [{\citenamefont {Pati}\ \emph {et~al.}(2000)\citenamefont {Pati},
  \citenamefont {Braunstein},\ and\ \citenamefont {Lloyd}}]{pati2000quantum}%
  \BibitemOpen
  \bibfield  {author} {\bibinfo {author} {\bibfnamefont {A.~K.}\ \bibnamefont
  {Pati}}, \bibinfo {author} {\bibfnamefont {S.~L.}\ \bibnamefont
  {Braunstein}},\ and\ \bibinfo {author} {\bibfnamefont {S.}~\bibnamefont
  {Lloyd}},\ }\bibfield  {title} {\bibinfo {title} {Quantum searching with
  continuous variables},\ }\href {https://arxiv.org/abs/quant-ph/0002082}
  {\bibfield  {journal} {\bibinfo  {journal} {arXiv:quant-ph/0002082}\ }
  (\bibinfo {year} {2000})}\BibitemShut {NoStop}%
\bibitem [{\citenamefont {Pati}\ and\ \citenamefont
  {Braunstein}(2003)}]{pati2003deutsch}%
  \BibitemOpen
  \bibfield  {author} {\bibinfo {author} {\bibfnamefont {A.~K.}\ \bibnamefont
  {Pati}}\ and\ \bibinfo {author} {\bibfnamefont {S.~L.}\ \bibnamefont
  {Braunstein}},\ }\bibfield  {title} {\bibinfo {title} {Deutsch-jozsa
  algorithm for continuous variables},\ }in\ \href@noop {} {\emph {\bibinfo
  {booktitle} {Quantum Information with Continuous Variables}}}\ (\bibinfo
  {publisher} {Springer},\ \bibinfo {year} {2003})\ pp.\ \bibinfo {pages}
  {31--36}\BibitemShut {NoStop}%
\bibitem [{\citenamefont {Braunstein}\ and\ \citenamefont {van
  Loock}(2005)}]{braunstein2005quantum}%
  \BibitemOpen
  \bibfield  {author} {\bibinfo {author} {\bibfnamefont {S.~L.}\ \bibnamefont
  {Braunstein}}\ and\ \bibinfo {author} {\bibfnamefont {P.}~\bibnamefont {van
  Loock}},\ }\bibfield  {title} {\bibinfo {title} {Quantum information with
  continuous variables},\ }\href {https://doi.org/10.1103/RevModPhys.77.513}
  {\bibfield  {journal} {\bibinfo  {journal} {Rev. Mod. Phys.}\ }\textbf
  {\bibinfo {volume} {77}},\ \bibinfo {pages} {513} (\bibinfo {year}
  {2005})}\BibitemShut {NoStop}%
\bibitem [{\citenamefont {Andersen}\ \emph {et~al.}(2010)\citenamefont
  {Andersen}, \citenamefont {Leuchs},\ and\ \citenamefont
  {Silberhorn}}]{andersen2010continuous}%
  \BibitemOpen
  \bibfield  {author} {\bibinfo {author} {\bibfnamefont {U.~L.}\ \bibnamefont
  {Andersen}}, \bibinfo {author} {\bibfnamefont {G.}~\bibnamefont {Leuchs}},\
  and\ \bibinfo {author} {\bibfnamefont {C.}~\bibnamefont {Silberhorn}},\
  }\bibfield  {title} {\bibinfo {title} {Continuous-variable quantum
  information processing},\ }\href {https://arxiv.org/abs/1008.3468} {\bibfield
   {journal} {\bibinfo  {journal} {Laser Photonics Rev.}\ }\textbf {\bibinfo
  {volume} {4}},\ \bibinfo {pages} {337} (\bibinfo {year} {2010})}\BibitemShut
  {NoStop}%
\bibitem [{\citenamefont {Jr}\ and\ \citenamefont
  {Kauffman}(2002)}]{lomonaco2002continuous}%
  \BibitemOpen
  \bibfield  {author} {\bibinfo {author} {\bibfnamefont {S.~J.~L.}\
  \bibnamefont {Jr}}\ and\ \bibinfo {author} {\bibfnamefont {L.~H.}\
  \bibnamefont {Kauffman}},\ }\bibfield  {title} {\bibinfo {title} {A
  continuous variable shor algorithm},\ }\href
  {https://arxiv.org/abs/quant-ph/0210141} {\bibfield  {journal} {\bibinfo
  {journal} {arXiv:quant-ph/0210141}\ } (\bibinfo {year} {2002})}\BibitemShut
  {NoStop}%
\bibitem [{\citenamefont {Zhong}\ \emph {et~al.}(2020)\citenamefont {Zhong},
  \citenamefont {Wang}, \citenamefont {Deng}, \citenamefont {Chen},
  \citenamefont {Peng}, \citenamefont {Luo}, \citenamefont {Qin}, \citenamefont
  {Wu}, \citenamefont {Ding}, \citenamefont {Hu} \emph
  {et~al.}}]{zhong2020quantum}%
  \BibitemOpen
  \bibfield  {author} {\bibinfo {author} {\bibfnamefont {H.-S.}\ \bibnamefont
  {Zhong}}, \bibinfo {author} {\bibfnamefont {H.}~\bibnamefont {Wang}},
  \bibinfo {author} {\bibfnamefont {Y.-H.}\ \bibnamefont {Deng}}, \bibinfo
  {author} {\bibfnamefont {M.-C.}\ \bibnamefont {Chen}}, \bibinfo {author}
  {\bibfnamefont {L.-C.}\ \bibnamefont {Peng}}, \bibinfo {author}
  {\bibfnamefont {Y.-H.}\ \bibnamefont {Luo}}, \bibinfo {author} {\bibfnamefont
  {J.}~\bibnamefont {Qin}}, \bibinfo {author} {\bibfnamefont {D.}~\bibnamefont
  {Wu}}, \bibinfo {author} {\bibfnamefont {X.}~\bibnamefont {Ding}}, \bibinfo
  {author} {\bibfnamefont {Y.}~\bibnamefont {Hu}}, \emph {et~al.},\ }\bibfield
  {title} {\bibinfo {title} {Quantum computational advantage using photons},\
  }\href {https://www.science.org/doi/10.1126/science.abe8770} {\bibfield
  {journal} {\bibinfo  {journal} {Science}\ }\textbf {\bibinfo {volume}
  {370}},\ \bibinfo {pages} {1460} (\bibinfo {year} {2020})}\BibitemShut
  {NoStop}%
\bibitem [{\citenamefont {Tillmann}\ \emph {et~al.}(2013)\citenamefont
  {Tillmann}, \citenamefont {Daki{\'c}}, \citenamefont {Heilmann},
  \citenamefont {Nolte}, \citenamefont {Szameit},\ and\ \citenamefont
  {Walther}}]{tillmann2013experimental}%
  \BibitemOpen
  \bibfield  {author} {\bibinfo {author} {\bibfnamefont {M.}~\bibnamefont
  {Tillmann}}, \bibinfo {author} {\bibfnamefont {B.}~\bibnamefont {Daki{\'c}}},
  \bibinfo {author} {\bibfnamefont {R.}~\bibnamefont {Heilmann}}, \bibinfo
  {author} {\bibfnamefont {S.}~\bibnamefont {Nolte}}, \bibinfo {author}
  {\bibfnamefont {A.}~\bibnamefont {Szameit}},\ and\ \bibinfo {author}
  {\bibfnamefont {P.}~\bibnamefont {Walther}},\ }\bibfield  {title} {\bibinfo
  {title} {Experimental boson sampling},\ }\href
  {https://www.nature.com/articles/nphoton.2013.102} {\bibfield  {journal}
  {\bibinfo  {journal} {Nat. Photonics}\ }\textbf {\bibinfo {volume} {7}},\
  \bibinfo {pages} {540} (\bibinfo {year} {2013})}\BibitemShut {NoStop}%
\bibitem [{\citenamefont {Marshall}\ \emph {et~al.}(2015)\citenamefont
  {Marshall}, \citenamefont {Pooser}, \citenamefont {Siopsis},\ and\
  \citenamefont {Weedbrook}}]{marshall2015quantum}%
  \BibitemOpen
  \bibfield  {author} {\bibinfo {author} {\bibfnamefont {K.}~\bibnamefont
  {Marshall}}, \bibinfo {author} {\bibfnamefont {R.}~\bibnamefont {Pooser}},
  \bibinfo {author} {\bibfnamefont {G.}~\bibnamefont {Siopsis}},\ and\ \bibinfo
  {author} {\bibfnamefont {C.}~\bibnamefont {Weedbrook}},\ }\bibfield  {title}
  {\bibinfo {title} {Quantum simulation of quantum field theory using
  continuous variables},\ }\href
  {https://journals.aps.org/pra/abstract/10.1103/PhysRevA.92.063825} {\bibfield
   {journal} {\bibinfo  {journal} {Phys. Rev. A}\ }\textbf {\bibinfo {volume}
  {92}},\ \bibinfo {pages} {063825} (\bibinfo {year} {2015})}\BibitemShut
  {NoStop}%
\bibitem [{\citenamefont {Yeter-Aydeniz}\ \emph {et~al.}(2022)\citenamefont
  {Yeter-Aydeniz}, \citenamefont {Moschandreou},\ and\ \citenamefont
  {Siopsis}}]{yeter2022quantum}%
  \BibitemOpen
  \bibfield  {author} {\bibinfo {author} {\bibfnamefont {K.}~\bibnamefont
  {Yeter-Aydeniz}}, \bibinfo {author} {\bibfnamefont {E.}~\bibnamefont
  {Moschandreou}},\ and\ \bibinfo {author} {\bibfnamefont {G.}~\bibnamefont
  {Siopsis}},\ }\bibfield  {title} {\bibinfo {title} {Quantum imaginary-time
  evolution algorithm for quantum field theories with continuous variables},\
  }\href {https://journals.aps.org/pra/abstract/10.1103/PhysRevA.105.012412}
  {\bibfield  {journal} {\bibinfo  {journal} {Phys. Rev. A}\ }\textbf {\bibinfo
  {volume} {105}},\ \bibinfo {pages} {012412} (\bibinfo {year}
  {2022})}\BibitemShut {NoStop}%
\bibitem [{\citenamefont {Culver}\ and\ \citenamefont
  {Schaich}(2022)}]{Culver:2021rxo}%
  \BibitemOpen
  \bibfield  {author} {\bibinfo {author} {\bibfnamefont {C.}~\bibnamefont
  {Culver}}\ and\ \bibinfo {author} {\bibfnamefont {D.}~\bibnamefont
  {Schaich}},\ }\bibfield  {title} {\bibinfo {title} {Quantum computing for
  lattice supersymmetry},\ }\href {https://arxiv.org/abs/2112.07651} {\bibfield
   {journal} {\bibinfo  {journal} {arXiv:2112.07651}\ } (\bibinfo {year}
  {2022})}\BibitemShut {NoStop}%
\bibitem [{\citenamefont {Schuld}\ and\ \citenamefont
  {Petruccione}(2021)}]{schuld2021machine}%
  \BibitemOpen
  \bibfield  {author} {\bibinfo {author} {\bibfnamefont {M.}~\bibnamefont
  {Schuld}}\ and\ \bibinfo {author} {\bibfnamefont {F.}~\bibnamefont
  {Petruccione}},\ }\href@noop {} {\emph {\bibinfo {title} {Machine learning
  with quantum computers}}}\ (\bibinfo  {publisher} {Springer},\ \bibinfo
  {year} {2021})\BibitemShut {NoStop}%
\bibitem [{\citenamefont {Killoran}\ \emph {et~al.}(2019)\citenamefont
  {Killoran}, \citenamefont {Bromley}, \citenamefont {Arrazola}, \citenamefont
  {Schuld}, \citenamefont {Quesada},\ and\ \citenamefont
  {Lloyd}}]{killoran2019continuous}%
  \BibitemOpen
  \bibfield  {author} {\bibinfo {author} {\bibfnamefont {N.}~\bibnamefont
  {Killoran}}, \bibinfo {author} {\bibfnamefont {T.~R.}\ \bibnamefont
  {Bromley}}, \bibinfo {author} {\bibfnamefont {J.~M.}\ \bibnamefont
  {Arrazola}}, \bibinfo {author} {\bibfnamefont {M.}~\bibnamefont {Schuld}},
  \bibinfo {author} {\bibfnamefont {N.}~\bibnamefont {Quesada}},\ and\ \bibinfo
  {author} {\bibfnamefont {S.}~\bibnamefont {Lloyd}},\ }\bibfield  {title}
  {\bibinfo {title} {Continuous-variable quantum neural networks},\ }\href
  {https://journals.aps.org/prresearch/abstract/10.1103/PhysRevResearch.1.033063}
  {\bibfield  {journal} {\bibinfo  {journal} {Phys. rev. res.}\ }\textbf
  {\bibinfo {volume} {1}},\ \bibinfo {pages} {033063} (\bibinfo {year}
  {2019})}\BibitemShut {NoStop}%
\bibitem [{\citenamefont {Arrazola}\ \emph {et~al.}(2019)\citenamefont
  {Arrazola}, \citenamefont {Bromley}, \citenamefont {Izaac}, \citenamefont
  {Myers}, \citenamefont {Br{\'a}dler},\ and\ \citenamefont
  {Killoran}}]{arrazola2019machine}%
  \BibitemOpen
  \bibfield  {author} {\bibinfo {author} {\bibfnamefont {J.~M.}\ \bibnamefont
  {Arrazola}}, \bibinfo {author} {\bibfnamefont {T.~R.}\ \bibnamefont
  {Bromley}}, \bibinfo {author} {\bibfnamefont {J.}~\bibnamefont {Izaac}},
  \bibinfo {author} {\bibfnamefont {C.~R.}\ \bibnamefont {Myers}}, \bibinfo
  {author} {\bibfnamefont {K.}~\bibnamefont {Br{\'a}dler}},\ and\ \bibinfo
  {author} {\bibfnamefont {N.}~\bibnamefont {Killoran}},\ }\bibfield  {title}
  {\bibinfo {title} {Machine learning method for state preparation and gate
  synthesis on photonic quantum computers},\ }\href
  {https://iopscience.iop.org/article/10.1088/2058-9565/aaf59e} {\bibfield
  {journal} {\bibinfo  {journal} {Quantum Sci. Technol.}\ }\textbf {\bibinfo
  {volume} {4}},\ \bibinfo {pages} {024004} (\bibinfo {year}
  {2019})}\BibitemShut {NoStop}%
\bibitem [{\citenamefont {Enomoto}\ \emph {et~al.}(2023)\citenamefont
  {Enomoto}, \citenamefont {Anai}, \citenamefont {Udagawa},\ and\ \citenamefont
  {Takeda}}]{enomoto2022continuous}%
  \BibitemOpen
  \bibfield  {author} {\bibinfo {author} {\bibfnamefont {Y.}~\bibnamefont
  {Enomoto}}, \bibinfo {author} {\bibfnamefont {K.}~\bibnamefont {Anai}},
  \bibinfo {author} {\bibfnamefont {K.}~\bibnamefont {Udagawa}},\ and\ \bibinfo
  {author} {\bibfnamefont {S.}~\bibnamefont {Takeda}},\ }\bibfield  {title}
  {\bibinfo {title} {Continuous-variable quantum approximate optimization on a
  programmable photonic quantum processor},\ }\href
  {https://doi.org/10.1103/PhysRevResearch.5.043005} {\bibfield  {journal}
  {\bibinfo  {journal} {Phys. Rev. Res.}\ }\textbf {\bibinfo {volume} {5}},\
  \bibinfo {pages} {043005} (\bibinfo {year} {2023})}\BibitemShut {NoStop}%
\bibitem [{\citenamefont {Bromley}\ \emph {et~al.}(2020)\citenamefont
  {Bromley}, \citenamefont {Arrazola}, \citenamefont {Jahangiri}, \citenamefont
  {Izaac}, \citenamefont {Quesada}, \citenamefont {Gran}, \citenamefont
  {Schuld}, \citenamefont {Swinarton}, \citenamefont {Zabaneh},\ and\
  \citenamefont {Killoran}}]{bromley2020applications}%
  \BibitemOpen
  \bibfield  {author} {\bibinfo {author} {\bibfnamefont {T.~R.}\ \bibnamefont
  {Bromley}}, \bibinfo {author} {\bibfnamefont {J.~M.}\ \bibnamefont
  {Arrazola}}, \bibinfo {author} {\bibfnamefont {S.}~\bibnamefont {Jahangiri}},
  \bibinfo {author} {\bibfnamefont {J.}~\bibnamefont {Izaac}}, \bibinfo
  {author} {\bibfnamefont {N.}~\bibnamefont {Quesada}}, \bibinfo {author}
  {\bibfnamefont {A.~D.}\ \bibnamefont {Gran}}, \bibinfo {author}
  {\bibfnamefont {M.}~\bibnamefont {Schuld}}, \bibinfo {author} {\bibfnamefont
  {J.}~\bibnamefont {Swinarton}}, \bibinfo {author} {\bibfnamefont
  {Z.}~\bibnamefont {Zabaneh}},\ and\ \bibinfo {author} {\bibfnamefont
  {N.}~\bibnamefont {Killoran}},\ }\bibfield  {title} {\bibinfo {title}
  {Applications of near-term photonic quantum computers: software and
  algorithms},\ }\href
  {https://iopscience.iop.org/article/10.1088/2058-9565/ab8504/meta} {\bibfield
   {journal} {\bibinfo  {journal} {Quantum Sci. Technol.}\ }\textbf {\bibinfo
  {volume} {5}},\ \bibinfo {pages} {034010} (\bibinfo {year}
  {2020})}\BibitemShut {NoStop}%
\bibitem [{\citenamefont {Herzberg}(1945)}]{herzberg1945molecular}%
  \BibitemOpen
  \bibfield  {author} {\bibinfo {author} {\bibfnamefont {G.}~\bibnamefont
  {Herzberg}},\ }\href@noop {} {\emph {\bibinfo {title} {Molecular spectra and
  molecular structure}}}\ (\bibinfo  {publisher} {D. van Nostrand},\ \bibinfo
  {address} {New York},\ \bibinfo {year} {1945})\BibitemShut {NoStop}%
\bibitem [{\citenamefont {Apanavicius}\ \emph {et~al.}(2021)\citenamefont
  {Apanavicius}, \citenamefont {Feng}, \citenamefont {Flores}, \citenamefont
  {Hassan},\ and\ \citenamefont {McGuigan}}]{apanavicius2021morse}%
  \BibitemOpen
  \bibfield  {author} {\bibinfo {author} {\bibfnamefont {J.}~\bibnamefont
  {Apanavicius}}, \bibinfo {author} {\bibfnamefont {Y.}~\bibnamefont {Feng}},
  \bibinfo {author} {\bibfnamefont {Y.}~\bibnamefont {Flores}}, \bibinfo
  {author} {\bibfnamefont {M.}~\bibnamefont {Hassan}},\ and\ \bibinfo {author}
  {\bibnamefont {McGuigan}},\ }\bibfield  {title} {\bibinfo {title} {Morse
  potential on a quantum computer for molecules and supersymmetric quantum
  mechanics},\ }\href {https://arxiv.org/abs/2102.05102} {\bibfield  {journal}
  {\bibinfo  {journal} {arXiv:2102.05102}\ } (\bibinfo {year}
  {2021})}\BibitemShut {NoStop}%
\bibitem [{\citenamefont {Roy}\ \emph {et~al.}(2006)\citenamefont {Roy},
  \citenamefont {Huang},\ and\ \citenamefont {Jary}}]{le2006accurate}%
  \BibitemOpen
  \bibfield  {author} {\bibinfo {author} {\bibfnamefont {R.~J.~L.}\
  \bibnamefont {Roy}}, \bibinfo {author} {\bibfnamefont {Y.}~\bibnamefont
  {Huang}},\ and\ \bibinfo {author} {\bibfnamefont {C.}~\bibnamefont {Jary}},\
  }\bibfield  {title} {\bibinfo {title} {An accurate analytic potential
  function for ground-state n2 from a direct-potential-fit analysis of
  spectroscopic data},\ }\href
  {https://pubs.aip.org/aip/jcp/article-abstract/125/16/164310/912126/An-accurate-analytic-potential-function-for-ground?redirectedFrom=fulltext}
  {\bibfield  {journal} {\bibinfo  {journal} {J. Chem. Phys.}\ }\textbf
  {\bibinfo {volume} {125}},\ \bibinfo {pages} {164310} (\bibinfo {year}
  {2006})}\BibitemShut {NoStop}%
\bibitem [{\citenamefont {Roy}\ and\ \citenamefont
  {Henderson}(2007)}]{roy2007new}%
  \BibitemOpen
  \bibfield  {author} {\bibinfo {author} {\bibfnamefont {R.~J.~L.}\
  \bibnamefont {Roy}}\ and\ \bibinfo {author} {\bibfnamefont {R.~D.~E.}\
  \bibnamefont {Henderson}},\ }\bibfield  {title} {\bibinfo {title} {A new
  potential function form incorporating extended long-range behaviour:
  application to ground-state ca2},\ }\href
  {https://www.tandfonline.com/doi/abs/10.1080/00268970701241656} {\bibfield
  {journal} {\bibinfo  {journal} {Mol. Phys.}\ }\textbf {\bibinfo {volume}
  {105}},\ \bibinfo {pages} {663} (\bibinfo {year} {2007})}\BibitemShut
  {NoStop}%
\bibitem [{\citenamefont {Chiu}\ \emph {et~al.}(2010)\citenamefont {Chiu},
  \citenamefont {Scott},\ and\ \citenamefont {Jakobsson}}]{chiumorse2010}%
  \BibitemOpen
  \bibfield  {author} {\bibinfo {author} {\bibfnamefont {S.-W.}\ \bibnamefont
  {Chiu}}, \bibinfo {author} {\bibfnamefont {H.~L.}\ \bibnamefont {Scott}},\
  and\ \bibinfo {author} {\bibfnamefont {E.}~\bibnamefont {Jakobsson}},\
  }\bibfield  {title} {\bibinfo {title} {A coarse-grained model based on morse
  potential for water and n-alkanes},\ }\href
  {https://doi.org/10.1021/ct900475p} {\bibfield  {journal} {\bibinfo
  {journal} {J. Chem. Theory Comput.}\ }\textbf {\bibinfo {volume} {6}},\
  \bibinfo {pages} {851} (\bibinfo {year} {2010})}\BibitemShut {NoStop}%
\bibitem [{\citenamefont {Jones}(2010)}]{Jones2010QuantumDO}%
  \BibitemOpen
  \bibfield  {author} {\bibinfo {author} {\bibfnamefont {A.}~\bibnamefont
  {Jones}},\ }\emph {\bibinfo {title} {Quantum drude oscillators for accurate
  many-body intermolecular forces}},\ \href@noop {} {Ph.D. thesis},\ \bibinfo
  {school} {University of Edinburgh} (\bibinfo {year} {2010})\BibitemShut
  {NoStop}%
\bibitem [{\citenamefont {Góger}\ \emph {et~al.}(2023)\citenamefont {Góger},
  \citenamefont {Khabibrakhmanov}, \citenamefont {Vaccarelli}, \citenamefont
  {Fedorov},\ and\ \citenamefont
  {Tkatchenko}}]{doi:10.1021/acs.jpclett.3c01221}%
  \BibitemOpen
  \bibfield  {author} {\bibinfo {author} {\bibfnamefont {S.}~\bibnamefont
  {Góger}}, \bibinfo {author} {\bibfnamefont {A.}~\bibnamefont
  {Khabibrakhmanov}}, \bibinfo {author} {\bibfnamefont {O.}~\bibnamefont
  {Vaccarelli}}, \bibinfo {author} {\bibfnamefont {D.~V.}\ \bibnamefont
  {Fedorov}},\ and\ \bibinfo {author} {\bibfnamefont {A.}~\bibnamefont
  {Tkatchenko}},\ }\bibfield  {title} {\bibinfo {title} {Optimized quantum
  drude oscillators for atomic and molecular response properties},\ }\href
  {https://doi.org/10.1021/acs.jpclett.3c01221} {\bibfield  {journal} {\bibinfo
   {journal} {J. Phys. Chem. Lett.}\ }\textbf {\bibinfo {volume} {14}},\
  \bibinfo {pages} {6217} (\bibinfo {year} {2023})}\BibitemShut {NoStop}%
\bibitem [{\citenamefont {Chabaud}\ \emph {et~al.}(2021)\citenamefont
  {Chabaud}, \citenamefont {Emeriau},\ and\ \citenamefont
  {Grosshans}}]{Chabaud:2021pnh}%
  \BibitemOpen
  \bibfield  {author} {\bibinfo {author} {\bibfnamefont {U.}~\bibnamefont
  {Chabaud}}, \bibinfo {author} {\bibfnamefont {P.-E.}\ \bibnamefont
  {Emeriau}},\ and\ \bibinfo {author} {\bibfnamefont {F.}~\bibnamefont
  {Grosshans}},\ }\bibfield  {title} {\bibinfo {title} {Witnessing wigner
  negativity},\ }\href {https://doi.org/10.22331/q-2021-06-08-471} {\bibfield
  {journal} {\bibinfo  {journal} {Quantum}\ }\textbf {\bibinfo {volume} {5}},\
  \bibinfo {pages} {471} (\bibinfo {year} {2021})}\BibitemShut {NoStop}%
\bibitem [{\citenamefont {Arkhipov}\ \emph {et~al.}(2018)\citenamefont
  {Arkhipov}, \citenamefont {Barasi{\'n}ski},\ and\ \citenamefont
  {Svozil{\'\i}k}}]{arkhipov2018negativity}%
  \BibitemOpen
  \bibfield  {author} {\bibinfo {author} {\bibfnamefont {I.~I.}\ \bibnamefont
  {Arkhipov}}, \bibinfo {author} {\bibfnamefont {A.}~\bibnamefont
  {Barasi{\'n}ski}},\ and\ \bibinfo {author} {\bibfnamefont {J.}~\bibnamefont
  {Svozil{\'\i}k}},\ }\bibfield  {title} {\bibinfo {title} {Negativity volume
  of the generalized wigner function as an entanglement witness for hybrid
  bipartite states},\ }\href
  {https://www.nature.com/articles/s41598-018-35330-6} {\bibfield  {journal}
  {\bibinfo  {journal} {Sci. Rep.}\ }\textbf {\bibinfo {volume} {8}},\ \bibinfo
  {pages} {16955} (\bibinfo {year} {2018})}\BibitemShut {NoStop}%
\bibitem [{\citenamefont {Lvovsky}\ and\ \citenamefont
  {Raymer}(2009)}]{Lvovsky:2009zz}%
  \BibitemOpen
  \bibfield  {author} {\bibinfo {author} {\bibfnamefont {A.~I.}\ \bibnamefont
  {Lvovsky}}\ and\ \bibinfo {author} {\bibfnamefont {M.~G.}\ \bibnamefont
  {Raymer}},\ }\bibfield  {title} {\bibinfo {title} {Continuous-variable
  optical quantum-state tomography},\ }\href
  {https://doi.org/10.1103/RevModPhys.81.299} {\bibfield  {journal} {\bibinfo
  {journal} {Rev. Mod. Phys.}\ }\textbf {\bibinfo {volume} {81}},\ \bibinfo
  {pages} {299} (\bibinfo {year} {2009})}\BibitemShut {NoStop}%
\bibitem [{\citenamefont {Puri}\ \emph {et~al.}(2017)\citenamefont {Puri},
  \citenamefont {Boutin},\ and\ \citenamefont {Blais}}]{puri2017engineering}%
  \BibitemOpen
  \bibfield  {author} {\bibinfo {author} {\bibfnamefont {S.}~\bibnamefont
  {Puri}}, \bibinfo {author} {\bibfnamefont {S.}~\bibnamefont {Boutin}},\ and\
  \bibinfo {author} {\bibfnamefont {A.}~\bibnamefont {Blais}},\ }\bibfield
  {title} {\bibinfo {title} {Engineering the quantum states of light in a
  kerr-nonlinear resonator by two-photon driving},\ }\href
  {https://www.nature.com/articles/s41534-017-0019-1} {\bibfield  {journal}
  {\bibinfo  {journal} {Npj Quantum Inf.}\ }\textbf {\bibinfo {volume} {3}},\
  \bibinfo {pages} {18} (\bibinfo {year} {2017})}\BibitemShut {NoStop}%
\bibitem [{\citenamefont {Grimm}\ \emph {et~al.}(2020)\citenamefont {Grimm},
  \citenamefont {Frattini}, \citenamefont {Puri}, \citenamefont {Mundhada},
  \citenamefont {Touzard}, \citenamefont {Mirrahimi}, \citenamefont {Girvin},
  \citenamefont {Shankar},\ and\ \citenamefont {Devoret}}]{grimm2019kerr}%
  \BibitemOpen
  \bibfield  {author} {\bibinfo {author} {\bibfnamefont {A.}~\bibnamefont
  {Grimm}}, \bibinfo {author} {\bibfnamefont {N.~E.}\ \bibnamefont {Frattini}},
  \bibinfo {author} {\bibfnamefont {S.}~\bibnamefont {Puri}}, \bibinfo {author}
  {\bibfnamefont {S.~O.}\ \bibnamefont {Mundhada}}, \bibinfo {author}
  {\bibfnamefont {S.}~\bibnamefont {Touzard}}, \bibinfo {author} {\bibfnamefont
  {M.}~\bibnamefont {Mirrahimi}}, \bibinfo {author} {\bibfnamefont {S.~M.}\
  \bibnamefont {Girvin}}, \bibinfo {author} {\bibfnamefont {S.}~\bibnamefont
  {Shankar}},\ and\ \bibinfo {author} {\bibfnamefont {M.~H.}\ \bibnamefont
  {Devoret}},\ }\bibfield  {title} {\bibinfo {title} {Stabilization and
  operation of a kerr-cat qubit},\ }\href
  {https://doi.org/10.1038/s41586-020-2587-z} {\bibfield  {journal} {\bibinfo
  {journal} {Nature}\ }\textbf {\bibinfo {volume} {584}},\ \bibinfo {pages}
  {205} (\bibinfo {year} {2020})}\BibitemShut {NoStop}%
\end{thebibliography}
\end{document}